\documentstyle[11pt,ophir]{article}

\begin{document}
\textheight 200mm
\topmargin -0.5cm
\parindent 0mm
\parskip 1em
\reversemarginpar
\newcommand{\ach}{\acosh}
\newcommand{\Cg}{\cosh^{2}{d^{\perp}_{g}(\zeta)}}
\newcommand{\oneoverCg}{\frac{1}{\Cg}}
\newcommand{\hp}{hph^{-1}}
\newcommand{\Chp}{\cosh^{2}{d^{\perp}_{\hp}(\zeta)}}
\newcommand{\smoothN}{\bar{N}}
\newcommand{\ch}{\cosh}
\newcommand{\sh}{\sinh}
\newcommand{\Cp}{\cosh^{2}{d^{\perp}_{p}(\zeta)}}
\newcommand{\di}{d_{\cal I}(\zeta)}
\newcommand{\dr}{d_{\cal R}(\zeta)}
\newcommand{\dg}{d_{g}(\zeta)}
\newcommand{\dq}{d_{q}(\zeta)}
\newcommand{\ds}{d_{s}(\zeta)}
\newcommand{\hq}{hqh^{-1}(\zeta)}
\newcommand{\hs}{hsh^{-1}(\zeta)}
\newcommand{\Chq}{\cosh^{2}{d^{\perp}_{\hq}(\zeta)}}
\newcommand{\Chs}{\cosh^{2}{d^{\perp}_{\hs}(\zeta)}}
\newcommand{\ChP}{\cosh^{2}{d^{\perp}_{h{\cal P}^rh^{-1}}(\zeta)}}
\newcommand{\Cq}{\cosh^{2}{d^{\perp}_{q}(\zeta)}}
\newcommand{\Cs}{\cosh^{2}{d^{\perp}_{s}(\zeta)}}
\def\po{\wp}
\pagestyle{empty}
\newpage
\renewcommand{\thefootnote}{*}
\pagestyle{empty}
\begin{center}
{\huge Exact Eigenfunctions of a Chaotic System} \\ \vspace{1cm}
{\Large Ophir M. Auslaender\footnote{Corresponding author. E-mail: ophir@physics.technion.ac.il; Tel:  972-4-829-3836; Fax: 972-4-822-1514}
 and Shmuel Fishman \\ Physics department - Technion, Haifa 32000, Israel \\}  \vspace{1cm}
\end{center}

\begin{abstract}
The interest in the properties of quantum systems, whose classical dynamics are chaotic, derives from their abundance in nature. The spectrum of such systems can be related, in the semiclassical approximation, to the unstable classical periodic orbits, through Gutzwiller's trace formula. The class of systems studied in this work, tiling billiards on the pseudo-sphere, is special in this correspondence being exact, via Selberg's trace formula.

In this work, an exact expression for Green's function and the eigenfunctions of tiling billiards on the pseudo-sphere, whose classical dynamics are chaotic, is derived. Green's function is shown to be equal to the quotient of two infinite sums over periodic orbits, where the denominator is the spectral determinant. Such a result is known to be true for typical chaotic systems, in the leading semiclassical approximation.  From the exact expression for Green's function, individual eigenfunctions can be identified.

In order to obtain a semiclassical approximation by finite series for the infinite sums encountered, resummation by analytic continuation in $\hbar$ was performed. The result, a semiclassical approximation, is similar to known results for eigenfunctions of typical chaotic systems. The lowest eigenfunctions of the Hamiltonian were calculated with the help of the resulting formulae, and compared with exact numerical results. A search for scars with the help of analytical and numerical methods failed to find evidence for their existence.
\end{abstract}
\vspace{3cm}
\newpage
\sf
\setcounter{page}{1}
\pagestyle{plain}
\renewcommand{\thefootnote}{\S\arabic{footnote}}
\setcounter{footnote}{0}
\setcounter{equation}{0}\def\theequation{\arabic{section}.\arabic{equation}}
\section{Introduction}\label{intro}
The spectrum of chaotic systems was studied extensively in recent years. Most of these studies are in the framework of the Gutzwiller trace formula \cite{Balian70,GutzwillerLH} for the density of states, that holds in the semiclassical limit, and was used successfully to analyze spectral correlations \cite{Wintgen86,Chang94,Fromhold94}. The main problem with it's application to the calculation of energy levels, is the fact that the Gutzwiller trace formula involves a non-absolutely convergent series. Resummation techniques enable to approximate the infinite sum by a finite one, and to obtain the spectrum in the semiclassical approximation \cite{Berry90,Berry92}. For billiards on surfaces of constant negative curvature, the density of states is given by the {\em exact} Selberg trace formula, that corresponds to the Gutzwiller trace formula. Since it is exact, it enables to distinguish between consequences of the semiclassical approximation of the Gutzwiller trace formula and those of other approximations, that are usually introduced, like the diagonal approximation or resummation.

Exploration of the eigenfunctions followed the studies of the spectrum. It is important because their properties determine matrix elements which control the nature of important physical phenomena, such as transport, dissociation of molecules, absorption and other physical processes where cross sections are important. Since the classical dynamics of many real systems are chaotic, the eigenfunctions of such systems are of special interest. Besides the physical motivation to study such eigenfunctions, they are also very interesting as they exhibit some surprising features. Prominent among these are scars, discovered by Heller \cite{Heller84} in numerical calculations. A scarred eigenfunction describes a state of the Hamiltonian of the system, whose probability density is unusually large in a narrow tube around an unstable classical periodic orbit. This effect was one of the factors that spurred the study of eigenfunctions of chaotic systems, which aimed at finding the explicit role classical periodic orbits play. Another important factor are mathematical theorems \cite{Shnirelman74} that state that the measure of scarred eigenfunctions (out of all eigenfunctions) is zero.

 In the present paper, an exact formula of eigenfunctions for some billiards on surfaces of constant negative curvature will be derived. It's main importance is for the understanding of the various approximations that were made in order to obtain the corresponding results in the semiclassical approximation for more realistic model systems.

Bogomolny \cite{Bogomolny88} studied the semiclassical limit of Green's function of chaotic systems. He showed that it has two kinds of contributions, the first being that of ``zero length orbits'' and the second consisting of a sum over all the classical periodic orbits. From this expression he obtained an expression for smoothed eigenfunctions, in which the smoothing is done over a small interval of energy. Because of this smoothing the study of individual eigenfunctions was prevented. Later Berry \cite{Berry89a} revealed the role periodic orbits play in the semiclassical limit of the spectral Wigner function. This function is closely related to Green's function, and is very suitable for semiclassical studies \cite{Berry77b}. Individual eigenfunctions are related to the poles of the spectral Wigner function. Agam and Fishman \cite{Agam93} cast the spectral Wigner function in a form that reveals it's poles. They showed that it can be written, in the semiclassical limit, as a ratio between two infinite sums over the classical periodic orbits. The denominator is the spectral determinant, whose zeros are the eigenenergies. This enables easy singling out of a specific eigenfunction. As the sums that appear in the numerator and the denominator were not absolutely convergent, they were treated by a resummation method previously invented by Berry and Keating \cite{Berry92} for the spectral determinant. Later on, Fishman \et \cite{Fishman96} showed that as Green's function is related to the solution of a Fredholm integral equation, it is given exactly by a ratio between two entire functions. The parameter in which the functions are entire is a book-keeping parameter that counts the number of times a surface of section through the system is crossed by Feynman paths. In addition, Fishman \et showed that invoking the semiclassical approximation, whose leading order arises from points near periodic orbits, Agam and Fishman's results are obtained. The term ``scars'' was quantified by Agam and Fishman \cite{Agam94a} who proposed to study the weight of the eigenfunction in a narrow tube around the classical periodic orbit. They found, for some model system, that the weight of scars in the eigenfunctions tends to zero in the semiclassical limit of vanishing wavelength. This decrease is very slow, in agreement with the observation of scars in numerical calculations.

In the present work, the diagonal part of Green's function of a billiard on a surface of constant negative curvature will be expressed as the ratio:
\beqa\label{ratio0}
G_E(\zeta,\zeta)=\frac{{\cal N}(\zeta,E)}{\Delta(E)}
~,
\eeqa
where $\Delta(E)$ is the spectral determinant (the precise meaning of $G_E(\zeta,\zeta)$ will be elucidated in \Sec{wave}). Exact expressions for ${\cal N}(\zeta,E)$ and $\Delta(E)$, in terms of classical orbits, will be obtained. These are similar, in their general structure, to expressions that were obtained in the framework of the semiclassical approximation, supplemented by other plausible approximations. It provides further evidence for the validity of the approximate formulas that were obtained for realistic systems.

The special chaotic systems that are investigated are billiards that are the primitive cells of lattices on a surface of constant negative curvature. This surface is called the pseudo-sphere. The reason for this choice is that the Green function of these systems is given exactly in terms of known contributions from classical orbits. In spite of the fact that these systems are chaotic, an exact expression for the Green function can be obtained by the method of images, since the surface is perfectly tiled by images of the billiards. 

The exploration of scars is of special interest for billiards on surfaces of constant negative curvature.  Rudnick and Sarnak \cite{Rudnick94} defined scars as eigenfunctions with weight that is divergent in the semiclassical limit. They proved that for a sub-class of tiling billiards on the pseudo-sphere (``Arithmetic Billiards'') there are no scars according to their definition. This is in agreement with the extrapolation of the semiclassical formula of Agam and Fishman. It does not make a clear statement about eigenfunctions for small, but finite, wavelengths. This issue was studied numerically for tiling billiards on surfaces of constant negative curvature \cite{Aurich91,Hejhal92,Aurich93}, and no scars were found. The quest for understanding of the difference between systems where scars are found and those where they are not was one of the motivations for developing analytical formulas for eigenfunctions for billiards on surfaces of constant negative curvature.

The outline of the paper follows. In the next two sections, results of previous research are reviewed~- \Sec{SC} is a brief summary of previous results concerning the semiclassical behaviour of typical chaotic systems. \Sec{neg} introduces the pseudo-sphere and tiling billiards on it, which are the specific chaotic systems analyzed in this work. In section \Sec{wave} the expression \Eq{ratio0} for Green's function is obtained. The exact eigenfunctions densities are it's residues. In order to truncate the infinite series that arise, Berry and Keating's resummation method \cite{Berry92} is used in section \Sec{sec_resum} to yield a semiclassical approximation to the exact expression. In section \Sec{Sec_numerics} some of the resummed expressions are compared to the exact eigenfunctions found numerically. In section \Sec{discuss}, the results of this work are compared with the ones found for general chaotic systems in the semiclassical approximation. The existence of scars is studied in section \Sec{Sec_PS_Yp}.

Many of the results in the paper rely on the review of Balazs and Voros \cite{Balazs86} and on the paper of Agam and Fishman \cite{Agam93}. Equations in these papers will be referred to as ``BV'' or ``AF'' followed by the equation number in the corresponding paper.

\setcounter{equation}{0}\def\theequation{\arabic{section}.\arabic{equation}}
\section{Semiclassics}\label{SC}
The starting point of the discussion is the Gutzwiller trace formula, which relates to the density of states of typical chaotic systems to the periodic orbits of the corresponding classical system. This formula is not convenient, as the density of states is a very singular function of energy. Therefore the spectral determinant, whose zeros lie on the spectrum of the Hamiltonian, is presented. In the semiclassical approximation the spectral determinant can also be written as an infinite sum over periodic orbits, that is obtained from the Gutzwiller series for the density of states. As similar sums appear many times in this work, Berry and Keating's method of resumming it to obtain truncated series is outlined. Finally, Agam and Fishman's results for eigenfunctions of typical chaotic systems are presented.

\subsection{Gutzwiller's trace formula}
The resolvent operator is given by:
\begin{equation}
\label{reso}
\hat{{\cal R}}(E+i\varepsilon)= \frac{1}{E+i\varepsilon -\hat{{\cal H}}}
~,
\end{equation}
where $\hat{\cal H}$ is the Hamiltonian operator. The retarded Green function is the representation of $\hat{{\cal R}}(E+i\varepsilon)$ in the coordinate-space basis:
\begin{equation}\label{generalGreen}
G_E^+({\bf r'},{\bf r})=\bra{{\bf r'}}\hat{{\cal R}}\ket{{\bf r}}=
\sum_{\alpha} \frac{\psi_{\alpha}^*({\bf r'})\psi_{\alpha}({\bf
r})}{E+i\eps-E_\alpha}{}_{\big|_{\eps\rightarrow0^+}} ~,
\end{equation}
and the density of states is related to it's trace:
\begin{equation}\label{TrR}
\rho(E)=\sum_\alpha\delta(E-E_\alpha)=-\frac{1}{\pi}\Im\left\{Tr\left[\hat{{\cal R}}(E+i\eps)\right]\right\}{}_{\big|_{\eps\rightarrow0^+}}
~.
\end{equation}
One would like to calculate the spectrum. Very often it is only possible in the leading order of the semiclassical approximation. For integrable systems, it can be done by the EBK method \cite{Tabor89}, that relies on the existence of $d$ ``good'' constants of motion, where $d$ is the number of degrees of freedom.

Gutzwiller \cite{GutzwillerLH} studied  $\rho(E)$ in the semiclassical limit for systems whose classical dynamics are very chaotic (hard chaos). In the semiclassical limit, the density of states can be divided into two parts with different analytic properties. The first of these is a smooth function of energy and the second is an oscillating function of it:
\begin{equation}
 \label{Semi_rho}
\rho(E) \sim \bar{\rho}(E)+\rho_{po}(E)\as\hbar0
~.
\end{equation}
The smooth part originates in the the system's short (non-periodic) classical orbits, and it's leading order is given by:
\beq
\label{TF}
\bar{\rho}(E)\sim\int \frac{d{\bf q}d{\bf p}}{h^d}\,\delta(E-{\cal H}\big({\bf q},{\bf p})\big)\as\hbar0
~.
\eeq
The oscillating part arises from the classical periodic orbits, and it's leading order  is given by the Gutzwiller trace formula \cite{GutzwillerLH}:
\begin{equation}
\label{Gutzwiller}
\rho_{po}(E)\sim\frac{1}{\pi\hbar}\Re\sum_{{\po}}T_{{\po}}\sum_{r=1}^{\infty}\frac{e^{\frac{i}{\hbar}\left(S_{{\po}}-i\gamma_{{\po}}\right)r}}{\sqrt{\ABS{\Det\left(M_{{\po}}^r-I\right)}}}\as\hbar0
~,
\end{equation}
where $\sum_{\po}$ denotes summation over all the primitive periodic orbits of the system, $T_{{\po}}$ denotes a primitive periodic orbit period, $S_{{\po}}$ it's action, $\gamma_{{\po}}$ the Maslov index associated with it and $M_{{\po}}$ denotes the monodromy matrix for motion perpendicular to it in it's vicinity in phase space.

As \Eq{Gutzwiller} stands it does not converge absolutely, due to the exponential proliferation of primitive periodic orbits with their length, typical of chaotic systems. Therefore, it is necessary to to treat Gutzwiller's trace formula further in order to obtain useful expressions. Such a treatment is presented in the next subsection.

\subsection{Spectral determinant and it's resummation}\label{sec_spectral}
For the calculation of the spectrum it is more convenient to study the spectral determinant. The semiclassical limit of this function was studied by Berry and Keating \cite{Berry90,Berry92}. In this section some of their results will be presented.

The regularized spectral determinant is defined by:
\beq\label{def_delta}
\Delta(E) = \prod_\alpha A(E,E_\alpha) (E-E_\alpha)
~,
\eeq
and it is convenient to study because it has zeros on the spectrum of the quantum Hamiltonian, instead of the poles the density of states \Eq{TrR} had. $A(E,E_\alpha)$ is just a regularizing function \cite{Voros87}.

Berry and Keating showed that the spectral determinant is given asymptotically by a dynamical Zeta function \cite{Berry90} (specializing for simplicity to two dimensions):
\beq
\label{Delta_prod}
\Delta(E)\sim B(E)e^{-i\pi\smoothN(E)}\prod_{{\po}}\prod_{m=0}^\infty\left\{1-e^{\frac i\hbar S_{{\po}}-i\gamma_{{\po}}}~e^{-\left(\half +m\right)u_{{\po}}}\right\}\as{\hbar}{0}
~,
\eeq
where $\smoothN(E)$ is the smooth part of the counting function of eigenenergies, given asymptotically Weyl's expansion in powers of $\hbar$ \cite{Balian70}.  $e^{\pm u_{{\po}}}$ are the eigenvalues of $M_{{\po}}$, and the $u_{{\po}} $'s are called instability exponents.The double product in \EQ{Delta_prod} can be written as an infinite sum:
\beq
\label{Delta_sum}
\Delta(E)\sim B(E)e^{-i\pi\smoothN(E)}\sum_\mu C_\mu e^{\frac i\hbar S_{\mu}}\as{\hbar}{0}
~,
\eeq
where $\mu$ denotes the list of composite periodic orbits ordered by length. In other words:
\bseq
\begin{eqnarray}
S_\mu&=&\sum_{\{j_\po\}_\mu}j_{\po}S_{\po}~~~~~j_{\po}=0,1,2,\ldots
\end{eqnarray}
and the coefficients $C_\mu$ are given by:
\beq
\label{Cmu_berry}
C_\mu=\prod_{\{\po\}_\mu}(-1)^{j_{\po}}e^{-ij_{\po}\gamma_{\po} }e^{-\qrt j_{\po}(j_{\po}-1)u_{\po}}\left|{{{\Prod_{j=1}^{j_{\po}}\Det{\left(M_{{\po}}^j-I\right)}}}}\right|^{-\half}
~.
\eeq
\eseq
\EQ{Delta_sum} suffers from the same convergence problems Gutzwiller's trace formula does. In order to treat them Berry and Keating introduced the resummation of \Eq{Delta_sum} by analytic continuation from the region of complex $\frac1\hbar$-plane where it converges. The method found to be effective made use of a Cauchy integral, and the result of the analytic continuation was found to be \cite{Berry92}:
\beq
\label{Delta_summed}
\Delta(E)\sim2\Re\Bigg\{\sum_\mu C_\mu(E)e^{-i\pi\smoothN(E)+\frac i\hbar S_\mu}\half \Erfc\left[\frac{\xi_\mu(\hbar,E)}{\sqrt{2\hbar Q^2(K,\hbar,E)}}\right]\Bigg\}\as{\hbar}{0}
~,
\eeq
where:
\bseq
\begin{eqnarray}
\label{spec_def_a}
\xi_\mu(\hbar,E)&=&S_\mu(E)-\frac{\Omega(E)}{h}
~,\\
\label{spec_def_b}
Q^2(K,\hbar,E)&=&K^2~+~i\frac{\Omega(E)}{h}
~.
\end{eqnarray}
\eseq
$K$ is a free tuning parameter chosen from the range $0<K<<\frac{d\Omega}{\Omega'}\sqrt{\frac{\hbar}{2}}\pi\overline{\rho}(E)$. This choice ensures that number of composite periodic orbits with actions in the cut-off region of the complementary error function is much smaller than the number of composite periodic orbits with action less than the center of the cut-off as $\hbar\rightarrow0$.

The subsequent terms in the asymptotic expansion for $\Delta(E)$, whose leading term is given by \Eq{Delta_summed}, are given by high order derivatives of the complementary error function. These derivatives are proportional to Gaussians centered around the complementary error function cut-off.

One observes that the resummation causes the effective truncation of the problematic sum in \Eq{Delta_sum}. The center of the cut-off is the composite periodic orbit whose period is half the Heisenberg time \cite{Berry92}:
\beq
{\cal T}_\mu^*(E)=\pi\hbar\bar{\rho}(E)
~.
\eeq

\subsection{Eigenfunctions of chaotic systems}\label{sec_wig}
Spectral Wigner's function is a function whose poles lie on the spectrum of the Hamiltonian, and whose residues are closely related to it's eigenstates. It's semiclassical behaviour was studied by Berry \cite{Berry89a,Berry89b}. Then, Agam and Fishman \cite{Agam93} showed, that in the semiclassical limit, it can be written as a ratio between two functions given by sums over periodic orbits. The denominator of the expression they found is the spectral determinant of the previous section. This allowed them to single out expressions for eigenfunctions.

Fishman \et \cite{Fishman96} were able to put the results of Agam and Fishman \cite{Agam93} on a more rigorous footing. They showed that the exact Green function is given by a ratio of two entire functions. Then they managed to show that in the semiclassical limit their result is identical with Agam and Fishman's result.

The method Fishman \et employed was Fredholm's method for the solution of certain integral equations. In this method the exact solution of the integral equation is given by the ratio between two functions. Each of the functions is given by an absolutely convergent series in powers of a transfer operator ($\hat T$) in a surface of section of the system.  Using a semiclassical expression for $\hat T$, the series then assumes the form of absolutely convergent sums over the classical orbits of the system. Then, using the same approximations used in \cite{Agam93}, results that concurred completely were obtained.

The expressions found by Agam and Fishman suffer from the same convergence problems \EQ{Gutzwiller} and \Eq{Delta_sum} do. This problem was solved for denominator by analytic continuation in the $\frac1\hbar$-plane by Berry and Keating \cite{Berry92}, as was mentioned in \Sec{sec_spectral}. The same treatment was implemented by Agam and Fishman for the numerator. Once this was done, they could present useful expressions for eigenfunction densities. The structure of these expressions is reminiscent of Gutzwiller's trace formula, for they have a smooth contribution and a periodic orbit contribution. For later reference only the latter contribution will be conveyed here (AF 3.64):
\beqa
\label{psi-O-msum}
\ABS{\Psi_\alpha({\bf q})}^2_{po} &\sim& \frac{4\pi}{h^{2} \Delta^{\prime}(E_\alpha)}\int_{-\infty}^{\infty}d{\bf P}\sum_{{\po}}\frac{\bar{\Delta}_i^\po\left( {\bf X},E_\alpha\right)}{\ABS{\dot q_{||}}}
\as\hbar0~,
\eeqa
where ${\bf\dot q_{||}}$ is the velocity along the primitive periodic orbit $\po$, ${\bf X}=(Q_{1,}\cdots Q_{d-1},P_{1},\cdots P_{d-1})$ are the phase space coordinates perpendicular to it, and:
\beqa
\label{delta_i_oded}
\bar{\Delta}_i^{\po}\left( {\bf X},E_\alpha\right) &\sim&\Im\Bigg\{\sum_{m=0}^{\infty}\sum_{\mu}C_{\mu}^{({\po},m)} g_m\left[\frac i\hbar{\bf{\widetilde X} R_{\po} X}\right]e^{\frac i\hbar{\bf{\widetilde X} R_{\po} X}} 
\times\Lf&&\times
e^{-i\pi\smoothN(E_{\alpha})+\frac i\hbar S_{\mu,\po}}\Erfc\left[\frac{\xi_{\mu,\po}(\hbar,E_{\alpha})+{\bf{\widetilde X} R_{\po} X}}{\sqrt{2\hbar Q^2(K, \hbar ,E_{\alpha})}}\right]\Bigg\}
~.
\eeqa
Confining the discussion to two dimensions, the argument of the complementary error function includes the function:
\beq\label{xi_mup_def}
{\xi}_{\mu,{\po}} (\hbar,E )={{\cal S}}_{\mu,{\po}}(E)~-~\frac{\Omega(E)}{2\pi\hbar}
~,
\eeq
and $Q^2(K, \hbar ,E_{\alpha})$ that was defined in \EQ{spec_def_b}. The coefficients $C_{\mu}^{({\po},m)}$ arise from expressing (AF 3.26):
\begin{eqnarray}\label{Deltapm_def}
\Delta^{({\po},m)}(E)=e^{-i\pi\smoothN (E)}&&\prod_{\{\po'\}\not=\po}\prod_{j=0}^{\infty} \left(1-e^{\frac{i}\hbar S_{\po'}-i\gamma_{\po'}}e^{-(\half+j)u_{\po'}}\right)
\times\Lf&&\times
\prod_{\stackrel{j=0}{j\not=m}}^{\infty} \left(1-e^{\frac{i}\hbar S_{{\po'}}-i\gamma_{\po'}}e^{-(\half+j)u_{{\po'}}}\right)\times e^{\frac{i}\hbar S_{{\po}}-i\gamma_{\po}}e^{-(\half+m)u_{\po}}
~,
\end{eqnarray}
as a sum over the composite orbits introduced in \Sec{sec_spectral}:
\begin{equation}\label{Deltapm_sum}
\Delta^{({\po},m)}(E)=\sum_{\mu}C_{\mu}^{({\po},m)} e^{-i\pi \smoothN(E) + 
\frac{i} {\hbar}{\cal S}_{\mu,{\po}}} 
~,
\end{equation}
\bseq
where, 
\begin{equation} 
\label{stam}
{\cal S}_{\mu,{\po}}={\cal S}_{\mu}+S_{{\po}}=(j_{\po}+1)S_{{\po}}+S_{\mu-{\po}}
~.
\end{equation}
$S_{\mu-{\po}}$ denotes the composite action $S_\mu$ after the contribution of the primitive periodic orbit $\po$ has been removed. The expression for the $C_{\mu}^{({\po},m)}$ given by (AF C.11) is identical with: 
\beq
\label{Cmupm}
C_{\mu}^{({\po},m)}~=~C_{\mu-{\po}}(-1)^{j_{\po}}e^{-\half(j_{\po}+1)u_{\po}-i(j_{\po}+1)\gamma_{\po}}\sum_{j=0}^{j_{\po}}(-1)^je^{-m(j+1)u_{\po}}d_{j_{\po}-j}\left(e^{-u_{\po}}\right)
~,
\eeq
where $C_{\mu-{\po}}$ is the coefficient of the composite orbit ${\cal S}_{\mu-{\po}}$ (see \EQ{stam}) of \Eq{Delta_sum}, and $d_{j}(x)$ is given by (AF C.2):
\beq\label{dj}
d_{j}(x)~=~\frac{x^{\half(j-1)j}}{\Prod_{r=1}^{j}(1-x^r)}
~.
\eeq
\eseq
The matrix ${\bf R}_{{\po}}({\bf x})$ is related to the eigenvectors of the monodromy matrix:
\begin{equation}
\label{Rp}
{\bf R}_{{\po}}({\bf x}) = \frac{1}{R_{{\po}}^{+}-R_{{\po}}^{-}}
\left(
 \begin{array}{cc}  -2R_{{\po}}^{+}R_{{\po}}^{-} & R_{{\po}}^{+}+R_{{\po}}^{-} \\
                     R_{{\po}}^{+}+R_{{\po}}^{-} & -2 \end{array} \right) ~~~~
\mbox{in which}~R_{{\po}}^\pm = \frac{v_{p2}^{\pm}}{v_{p1}^{\pm}}
~.
\end{equation}
The function $g_m\left[x\right]$ appears through an expansion in powers of $e^{-u_\po}$ (AF 3.11 \& 3.13):
\beq 
\label{taylor}
\frac{\exp\left[  
\frac{i}{\hbar} \widetilde{{\bf X}} J \frac
{M^r_{\po}-I}{M^r_{\po}+I} {\bf X}\right]}{\sqrt{\det (M^r_{\po}+I)}}~=~\sum_{m=0}^\infty \!g_m\!\!\left[\frac i\hbar{\bf{\widetilde X} R_{\po} X}\right]e^{\frac i\hbar{\bf{\widetilde X} R_{\po} X}}~e^{-r\left(\half+m\right)u_{\po}}
~,
\eeq
where $r$ was used for repetitions.

In \App{Msum} it is shown that:
\beqa
\label{delta_i_ophir}
\bar{\Delta}_i^{\po}\left( {\bf X},E_\alpha\right) &\sim&\Im\Bigg\{\sum_{\mu} \sum_{j=0}^{j_{\po}}
\frac{\bar{C}_\mu^{({\po},j)}}{2\cosh{\frac{j+1}2u_{\po}}} 
e^{  \frac{i}{\hbar} {\bf{\widetilde X} R_{\po} X}~\tanh{\frac{j+1}2u_{{\po}}}} 
\times\\&&\vspace{-2cm}\times 
e^{-i\pi\smoothN(E_{\alpha})+\frac{i}{\hbar}{{\cal S}}_{\mu,{\po}}(E_{\alpha})}
\Erfc\left[ \frac{{\xi}_{\mu,{\po}} (\hbar,E_{\alpha} ) +{\bf{\widetilde X} R_{\po} X}} {\sqrt{2\hbar Q^2(K, \hbar ,E_{\alpha})}} \right]\Bigg\}
\as{\hbar}{0}
~,\nonumber
\eeqa
in which another definition was introduced:
\beq
\label{def_Cmupj}
\bar{C}_\mu^{({\po},j)} \equiv C_{\mu-{\po}} e^{-i(j_{\po}+1)\gamma_{\po}} (-1)^{j_{\po}-j} e^{-\half(j_{\po}-j)u_{\po}} d_{j_{\po}-j}\left(e^{-u_{\po}}\right)
~.
\eeq
\newpage
Finally, the integral in \Eq{psi-O-msum} can be performed using the stationary phase approximation. The result is (AF 4.3):
\beqa
\label{psi-O}
\ABS{\Psi_\alpha({\bf q})}^2_{po} &\sim& \frac{4\pi}{h^{2} \Delta^{\prime}(E_\alpha)}\Im\Bigg\{\sum_{\po}\frac{1}{\ABS{\dot q_{||}}}\sqrt{\frac{\hbar\pi}{2i} \half\left(R^+_{\po}-R^-_{\po}\right)}\sum_\mu \sum_{j=0}^{j_{\po}}
\frac{e^{  \frac{i}{\hbar} \half\left[R^+_{\po}-R^-_{\po}\right]q_{\perp}^2\tanh{\frac{j+1}2u_{{\po}}}}}{\sqrt{\sinh{(j+1)u_{\po}}}}
\bar{C}_\mu^{({\po},j)}
 \times\Lf
&&\!\!\!\!\!\!\!\!\!\!\!\!\!\!\!\times e^{-i\pi\smoothN(E_{\alpha})+\frac{i}{\hbar}{{\cal S}}_{\mu,{\po}}(E_{\alpha})}\Erfc\left[ \frac{{\xi}_{\mu,{\po}} (\hbar,E_{\alpha} ) +\half\left[R^+_{\po}-R^-_{\po}\right]q_{\perp}^2} {\sqrt{2\hbar Q^2(K, \hbar ,E_{\alpha})}}\Bigg\} \right]\as{\hbar}{0}
~.
\eeqa

\setcounter{equation}{0}\def\theequation{\arabic{section}.\arabic{equation}}
\section{Tiling billiards on the pseudo-sphere}\label{neg}
The systems considered in this work are tiling billiards on the pseudo-sphere. There is extensive mathematical literature concerning these systems. Various aspects of geometry can be found in \cite{Beardon83}, while many of the results about quantum mechanical properties of these systems, to be conveyed in the next subsections, can be found in \cite{Selberg56,Hejhal76,Hejhal83}. Useful reviews are \cite{Balazs86,Bogomolny96}

The pseudo-sphere is a surface of constant negative Gaussian curvature of $-\frac1{R^2}$. Setting $R=1$, all distances are measured in units of $R$. Because the surface is two dimensional, it can be projected onto the the complex plane, so that the two dimensional coordinates are given by complex numbers. Two very useful projections are the Poincar\'e disk, in which the surface is projected onto the interior of the unit disk, and the Poincar\'e half plane, in which it is projected onto the upper half of the complex plane. In both of this models the geodesics on the surface turn out to be arcs of circles perpendicular to the boundary of the model. 

The invariant measures in both models are:
\bseq
\begin{eqnarray}
	\label{Disk_dmu}  d\mu(z)&=&\frac{4r~drd\phi}{\left(1-r^2\right)^2}~~~\mbox{for the disk,}\\
	\label{plane_dmu}  d\mu(\zeta)&=&\frac{dxdy}{y^2}~~~\mbox{for the half-plane,}
\end{eqnarray}
\eseq
where $z=re^{i\phi}$ are disk coordinates, and $\zeta=x+iy$ are half-plane coordinates. The distance between two points is:
\bseq
\begin{eqnarray}
\label{diskDist}
\ch{d_{z,z'}}&=&1+\frac{2\left|z'-z\right|^2}{\left(1-|z|^2\right)\left(1-|z'|^2\right)}~~~\mbox{for the disk,}\\
\label{planeDist}
\cosh d_{\zeta,\zeta'} &=& 1 + \half\frac{\left|\zeta'-\zeta\right|^2}{y'~y}~~~\mbox{for the half-plane.}
\end{eqnarray}
\eseq

There are several kinds of isometries that are going to be encountered in this paper:
\begin{enumerate}
\item \underline{Even boosts}: these isometries are the analogues of Euclidean translations. They have two fixed points at infinity, which lie on the boundary of both of the Poincar\'e models. Any two points on the surface are connected by a unique geodesic. The unique geodesic that connects these fixed points is called the boost's {\em invariant geodesic}.
\item \underline{Rotations}: these isometries are the analogues of Euclidean rotations. They have one fixed point which does not lie at infinity.
\item \underline{Inversions}: reflections across the invariant geodesics of boosts.
\item \underline{Odd boosts}: an even boost followed by an inversion through it's invariant geodesic.
\end{enumerate}
One of the important facts about the pseudo-sphere is that the isometries listed above, are given by M\"obius transformations in the Poincar\'e disk and half-plane:
\beq
\zeta'=g(\zeta)=\frac{a\zeta+b}{c\zeta+d}~~~~ad-bc=1,~a,b,c,d\in\real~,
\eeq
in the case of even boosts and rotations (in the half-plane), and
\beq
\zeta'=g(\zeta)=\frac{a\zeta^\ast+b}{c\zeta^\ast+d}~~~~ad-bc=-1,~a,b,c,d\in\real~,
\eeq
for inversions and odd boosts (in the half-plane). This means that an isometry that is obtained by a sequence of isometries, is the product of the appropriate matrices.

Many statements about the invariant geodesics of even and odd boosts are going to be made in the following discussion. They are usually easier to consider in the half-plane model, because in it the positive real axis is a geodesic, therefore it is always possible to chose coordinates such that it is the invariant geodesic of the boost under consideration. In this frame of reference the matrix that corresponds to a boost is diagonal:
\begin{equation}
\label{diagP}
g=\left(\begin{array}{cc} \lambda&0\\
			        0&\frac{\pm 1}{\lambda} \end{array}\right)~~~~~~\left.\begin{array}{c} +~~\mbox{even boost}\\-~~\mbox{odd boost}\end{array}\right.
\end{equation}

As the curvature is negative, the flow upon the pseudo-sphere is hyperbolic. This means that adjacent trajectories diverge exponentially fast from each other. In spite of this, motion on the whole surface is integrable \cite{Balazs86}. According to \cite{Ornstein73}, hyperbolic flow on a compact surface is chaotic. Therefore, motion in any compact surface on the pseudo-sphere is assured to be chaotic. One such choice are tiling billiards. Such billiards are the primitive cells of lattices of geodesics on the pseudo-sphere, as can be seen in \Fig{tile}. The word ``billiard'' refers to an enclosure on the pseudo-sphere with any boundary conditions. In this paper two cases are going to be concentrate upon, namely periodic and Dirichlet boundary conditions.

The situation presented above is very different from the one arising in the Euclidean plane. In that case trajectories diverge only linearly on the whole surface, so that in order to ensure the chaotic nature of the motion complicated boundaries have to be chosen to introduce the necessary exponential divergence. 

Having a billiard that is a primitive cell of a lattice of geodesics, means that there is an underlaying discrete group of isometries, whose fundamental domain is the primitive cell. A fundamental domain of such a discrete group is a domain, that under the action of all of the different members of the group tiles the whole surface perfectly. The simplest group consists of only even boosts (in addition to the identity), the fundamental domain of which is a polygon with periodic boundary conditions. If the fundamental domain is to be a billiard with hard walls, than the group consists of even boosts, odd boosts, reflections and rotations.

An interesting feature of tiling billiards on the pseudo-sphere is that their periodic orbits are closely related to the boosts of the underlying group. This situation also exists in tiling billiards on the Euclidean plane (with translations instead of boosts). The important difference between these systems is that in the latter case the underlying group is Abelian, whereas in the former it isn't. In addition, on the Euclidean plane there are continuous families of periodic orbits, whereas on the pseudo-sphere all periodic orbits are isolated.

Consider free motion inside the fundamental domain, the billiard. Since the motion is free until the boundary is hit, it is on geodesics. Once one of the edges is hit the trajectory proceeds according to the boundary conditions. Again, the simplest case is that of periodic boundary conditions, in which the trajectory re-enters from the edge identified with the one it hit, on a new geodesic. A more mathematically complicated case is the case of reflections, in which the trajectory is reflected specularly and continues on the appropriate geodesic.

There is another way to describe the motion, in which the fact that all the different copies of the billiard, under the discrete group of isometries, tile the whole surface. Beginning on a geodesic segment in the billiard and travelling on it with constant velocity, eventually one of the edges is hit. Then, instead of staying in the billiard on a different geodesic, the original geodesic is continued into the billiard copy across the edge. If the boundary conditions were periodic this copy is identical to the original billiard, and if the edge is a hard wall, the copy's orientation is reversed with respect to the original billiard, since it is obtained by reflection.

Now, the next segment of the trajectory is on the original geodesic, but in a different copy of the billiard. All this process will have to be repeated once another edge is hit. Continuing the construction {\em ad infinitum}, the trajectory is described by the collection of the different copies of the billiard through which the original geodesic passes.

Recapitulating, instead of describing the classical trajectory in the usual way, as a collection of geodesic segments, all of which lay inside a limited area of the surface, it can be described with one geodesic that enters and exits copies of the original billiard.

The above procedure is standard practice in polygonal billiards (see for example \cite{Richens81}). The next step's presentation is due to Bogomolny \et \cite{Bogomolny97}:
\begin{enumerate}
	\item Consider one of the geodesics described above, that enters and exits many tiles of the tessellation;
	\item Each tile along it is labeled by the group isometry with which it was obtained from the fundamental domain: $\left\{I,g_1,g_2,p,\ldots\right\}$ ($I$ is the first segment, the one in the original copy);
	\item To reconstruct the ''physical'' trajectory, the geodesic must be folded back into the original fundamental domain. This is done by copying each geodesic segment by the isometry inverse to the one that created the copy it is in. These isometries are: $\left\{I,g_1^{-1},g_2^{-1},p^{-1},\ldots\right\}$;
	\item After this operation most of the geodesic segments will be dispersed around in a complicated manner. But, for special trajectories, one of the segments (the one that was in the copy that was suggestively labeled $p$) will match the first segment {\em exactly} in location {\em and} in orientation. In this way:
	\begin{quotation}\noindent an isometry that copies a geodesic segment onto a segment with identical orientation of the same  geodesic has been found. Therefore:
		\begin{enumerate}
			\item The geodesic segments constitute a periodic orbit;
			\item This geodesic is the invariant geodesic of $p$;
			\item On the pseudo-sphere $p$ has to be a boost, for it has an invariant  geodesic. (It can't be a pure reflection because the orientation is the same.) 
		\end{enumerate}
	\end{quotation} 
\end{enumerate}
The arguments given here are applicable to tiling billiards on the plane too, where the role of the boosts is taken by simple translations. The qualitative difference is that on the plane, periodic orbits are not isolated, whereas on the pseudo-sphere they are, due to the exponential divergence of close trajectories.

The distance a point is transfered by a boost  is closely related to the it's distance from the boost's invariant geodesic. If $g$ is an even boost \cite{Beardon83}:
\bseq
\beq
\label{beardoneven}
\sinh{\half d_g(\zeta)}=\cosh{d^{\perp}_{g}(\zeta)}~\sinh{\half L_g}~,
\eeq
where $d_g(\zeta)\equiv d_{\zeta,g(\zeta)}$ is the distance of a point from it's image under $g$, and $d^{\perp}_{g}(\zeta)$ is the distance of the point $\zeta$ from the invariant geodesic of $g$ measured along a perpendicular geodesic (see \Fig{fig_perp}). A similar relation can easily be obtained for odd boosts:
\beq
\label{beardonodd}
\cosh{\half d_g(\zeta)}=\cosh{d^{\perp}_{g}(\zeta)}~\cosh{\half L_g}
~.
\eeq
\eseq

\subsection{Green's function for fundamental domain billiards}
The Green function is calculated exactly by the method of images, from the free Green function, by virtue of the underlying discrete group of isometries. In half-plane coordinates, Schr\"odinger's equation reads \cite{Balazs86}:
\begin{eqnarray}
\label{SE}
-\frac{\hbar^2}{2 m}~ y^2~\left(\Npartl{x}{2}+\Npartl{y}{2}\right)\Psi=E\Psi
~.
\end{eqnarray}
Distance is measured in units of $R$, and from this point on, energy will be measured in units of $\frac{\hbar^2}{2mR^2}$ so that = if $m=\half$ then $\hbar=1$ . In these units, it turns out that the free retarded Green function is given by \cite{Balazs86}:
\begin{equation}
\label{freegreen}
G_E^{0}(\zeta,\zeta')=\frac{-1}{2\pi}~Q_{-\half - ik}\left[\cosh{d_{\zeta,\zeta'}}\right]
~,
\end{equation}
where $E=k^2+\qrt$ and $Q_l[x]$ is Legendre's function of the second kind. 

The billiards under consideration are the fundamental domains of groups of isometries. Therefore, all the copies of the billiard tile the surface perfectly. In addition, as all the copies are identical, a solution of Schr\"odinger's equation in one, solves it in all the other tiles, but for a coordinate transformation. This means that Green's function has to be invariant under the action of the group isometries ($g_i\in\Gamma$). A simple way to obtain such a function is to use the method of images and free Green's function:
\begin{eqnarray}
\label{Green_group}
G_E(\zeta,\zeta')=\sum_{g\in\Gamma}\chi_{g}G_E^0(\zeta,g(\zeta'))
~.
\end{eqnarray}
where $\chi_{g}=(\pm1)^{parity~of~g}$ ($+1$ is the choice for Neumann and periodic boundary conditions and $-1$ for the Dirichlet case).

Now the {\em exact} retarded Green function for fundamental domain billiards can be written using \Eq{freegreen}:
\begin{eqnarray}
\label{Green_exact}
G_E(\zeta,\zeta')=\frac{-1}{2\pi}\sum_{g\in\Gamma}\chi_{g}Q_{-\half - ik}\left[\cosh{d_{\zeta,g(\zeta')}}\right]
~.
\end{eqnarray}
\EQ{Green_exact} has an intuitive explanation. It gives the Green function between $\zeta$ and $\zeta'$ as a sum over all the classical orbits between them. To see this one should describe a complicated orbit between these two points using the scheme described already in the previous subsection. The distance between $\zeta$ and one of the copies $g(\zeta')$ is measured along the unique geodesic that connects them. As $g(\zeta')$ is completely equivalent to $\zeta'$, this geodesic is an (unfolded) classical orbit between $\zeta$ and $\zeta'$. This is fundamentally different from other quantum mechanical problems, where all Feynman paths are required for the calculation of Green's function. This property, that holds when tiling is possible, makes the semiclassical approximation exact for the problems studied here.

The sum over the group members of \Eq{Green_exact} can be arranged in a suggestive manner. Any group can be divided into {\em conjugacy classes} \cite{Hamermesh62}. A conjugacy class is defined as the collection of all different group members that are similar to one another. All the members of a conjugacy class can be generated from one representative via:
\begin{equation}
\label{sim}
g'=h^{-1}gh~~~h\in\Gamma/\Gamma_g
~,
\end{equation}
where $\Gamma/\Gamma_g$ is the sub-group of $\Gamma$, whose members are unity and the members that do not commute with $g$. The reason for excluding $\Gamma_g$ (the sub-group of the group members that commute the $g$) is that each member of a conjugacy class should be accounted for only once.

It may be shown that there is a one-to-one correspondence between the conjugacy classes of boosts and the periodic orbits \cite{Selberg56,Balazs86}. This means that the list of boost conjugacy classes is the list of periodic orbits. Therefore, it has the form:
\begin{eqnarray*}
\{\dots,p_i^r,\dots\}~,
\end{eqnarray*}
where $r=1,2,3,\dots$ are repetitions and $\{p_i\}$ is the list of primitive periodic orbits of the billiard. Note that $p$ and $p^{-1}$ can only be conjugate if there are reflections in the group, {\em i.e} if the periodic orbit is self retracing.

The exact retarded Green function is given by \Eq{Green_exact}. Rearranging the sum over boosts in complete boost conjugacy classes, one obtains:
\begin{eqnarray}
\label{Green_exact2}
G_E(\zeta,\zeta')&=&\frac{-1}{2\pi}\hspace{-0.6cm}\sum_{\stackrel{g}{\mbox{\scriptsize{not a boost}}}}\hspace{-0.6cm}\chi_{g}Q_{-\half - ik}\left[\cosh d_{\zeta,g(\zeta')}\right]~+
\Lf&&\hspace{3cm}+~
\frac{-1}{2\pi}\sum_{\{\po\}}\sum_{r=1}^{\infty}\sum_{h\in\Gamma/\Gamma_\po}\chi_{\po^r}Q_{-\half - ik}\left[\cosh d_{\zeta,h^{-1}\po^rh(\zeta')}\right]~,
\end{eqnarray}
where $\sum_{\{\po\}}\sum_{r=1}^{\infty}$ is the sum over all the different boost conjugacy classes, and $\sum_{h\in\Gamma/\Gamma_\po}$ is the sum over the different members of each class.

For a billiard with periodic boundary conditions, the group of isometries, whose fundamental domain is the billiard, contains only the identity and even boosts. Retarded Green's function is given, in this case, by:
\begin{eqnarray}
\label{Green_PBC}
G_E(\zeta,\zeta')&=&\frac{-1}{2\pi}Q_{-\half - ik}\left[\cosh d_{\zeta,\zeta'}\right]~+
\Lf&&\hspace{3cm}+~
\frac{-1}{2\pi}\sum_{\{p\}}\sum_{r=1}^{\infty}\sum_{h\in\Gamma/\Gamma_p}Q_{-\half - ik}\left[\cosh d_{\zeta,h^{-1}p^rh(\zeta')}\right]~,
\end{eqnarray}
where $\sum_{\{p\}}\sum_{r=1}^{\infty}$ is the sum over all the different even boost conjugacy classes.

For a billiard with hard walls, the group whose fundamental domain it is, is generated by reflections at the edges. This means that it will include the following types of isometries:
\begin{enumerate}
	\item The identity $I$.
	\item Inversions $\{{\cal I}\}$ -  reflections through the edges of all the tiles of the tessellation.
	\item Rotations $\{{\cal R}\}$ - rotations around all of the vertices of all of the tiles in the tessellation.
	\item Singular boosts - hyperbolic isometries that commute with some reflection. They correspond to periodic orbits that run along edges of the billiard. These periodic orbits are invariant to reflections across those edges, so singular boosts always come in pairs of odd and even.
	\item Odd boosts - correspond to periodic orbits that hit an odd number of edges of the billiard.
	\item Even boosts - correspond to periodic orbits that hit an even number of edges of the billiard. 
\end{enumerate}
As before, the boosts can be divided into conjugacy classes, whose list is the list of periodic orbits, which are repetitions of primitive periodic orbits. Since there are inversions in the group, there are two types of primitive periodic orbits: those that bounce an even number of times and those that bounce an odd number of times, during one period. The former will be denoted by $\{p_i\}$, and the latter by $\{q_i\}$.

It is convenient to separate the singular boosts from the rest. The previous argumentation about hyperbolic isometries applies to them, so they too can be ordered according to conjugacy classes. One of the peculiarities about the orbits that correspond to their conjugacy classes, is that for every even primitive singular orbit, there is an odd one. This is why these conjugacy classes can be ordered in pairs: $\{s\}$ and $\{{\cal I}_ss\}$, where ${\cal I}_s$ is an inversion that commutes with $s$.
\newpage
Now \Eq{Green_exact2} can used again to write the exact retarded Green function of the problem:
\begin{eqnarray}\label{Green_hardwall}
G_E(\zeta,\zeta')&=&\frac{-1}{2\pi}\Bigg\{Q_{-\half-ik}\left[d_{\zeta,\zeta'}\right]+\sum_{\{\cal R\}}Q_{-\half-ik}\left[\cosh d_{\zeta,{\cal R}(\zeta')}\right]+\epsilon\sum_{\{\cal I\}}Q_{-\half-ik}\left[\cosh d_{\zeta,{\cal I}(\zeta')}\right]+
\Lf&&\!\!\!\!\!\!\!\!\!\!\!\!\!\!\!\!\!\!\!\!\!\!\!\!\!\!\!
+\sum_{\{p\}}\sum_{r=1}^\infty\sum_{h\in\Gamma/\Gamma_p}Q_{-\half-ik}\left[\cosh d_{\zeta,h^{-1}p^rh(\zeta')}\right]
+
\sum_{\{q\}}\sum_{r=1}^\infty\epsilon^r\!\!\!\sum_{h\in\Gamma/\Gamma_q}Q_{-\half-ik}\left[\cosh d_{\zeta,h^{-1}q^rh(\zeta')}\right]+
\Lf&&\!\!\!\!\!\!\!\!\!\!\!\!\!\!\!\!\!\!\!\!\!\!\!\!\!\!\!
+\sum_{\{s\}}\sum_{r=1}^\infty\sum_{h\in\Gamma/\Gamma_s}\left[Q_{-\half-ik}\left[\cosh d_{\zeta,h^{-1}s^rh(\zeta')}\right]+\epsilon Q_{-\half-ik}\left[\cosh d_{\zeta,h^{-1}{\cal I}_ss^rh(\zeta')}\right]\right]\Bigg\}
~,
\end{eqnarray}
where $\epsilon=-1$ for Dirichlet boundary conditions, and $\epsilon=+1$ for Neumann boundary conditions. In this work only the former will be studied, because the generalization for the latter is straightforward. The term ``hard wall billiard'' will be used from this point forth to describe a ``Dirichlet boundary conditions' billiard''.

\subsection{The density of states and Selberg's trace formula}\label{sec_density}
The density of states can be calculated exactly by tracing Green's function. For simplicity, a billiard with periodic boundary conditions will be considered initially. Tracing \Eq{Green_PBC} yields the density of states that takes the form \cite{Balazs86}:
\begin{eqnarray}\label{Selberg}
{\rho}(E)=\bar{\rho}(E)~+~\rho_{po}(E)=\frac{{\cal A}}{4\pi}~\tanh{\pi k}+\Re\left\{\sum_{\{p\}}\frac{L_p}{4\pi k}\sum_{r=1}^\infty~\frac{e^{ikrL_p}}{\sinh{\frac{r}{2}L_p}}\right\}
~,
\end{eqnarray}
where ${\cal A}$ is the area of the billiard. \EQ{Selberg} is the Selberg trace formula \cite{Selberg56}.

Using \EQ{Selberg}, one can derive Weyl's asymptotic expansion for the counting function of eigenvalues (BV VII.36):
\begin{equation}
\label{weyl_a}
\smoothN(E)\sim \frac{{\cal A}}{4\pi}\left(E-\frac{1}{3}\right)~~~\mbox{as}~~E\rightarrow\infty
\end{equation}
It turns out that all other terms in the series are exponentially small as $E\rightarrow\infty$.

The computation of the density of states in hard wall billiards is very similar in principle, so it will not be presented. A detailed computation appears in \cite{Balazs86}. The result of tracing \Eq{Green_hardwall} is:
\beq
\label{SelbergBill}
{\rho}(E)=\overline{\rho}(E)~+~\rho_{po}(E)
~,
\eeq
where (BV M.14):
\begin{eqnarray}\label{SelbergBillSmooth}
\bar{\rho}(E)&\equiv&
	\Bigg\{\frac{{\cal A}}{4\pi}~\tanh{\pi k}~-~\frac{{\cal L}}{8\pi k}
+\Lf&&+
\frac{1}{8\pi k}\sum_{\{\frac{\pi}{m_r}\}}\frac1{m_r}\int_{-\infty}^{\infty}\frac{dx \cos{kx}}{\sinh \frac{x}{2}}\left(m_{r}\coth\frac{m_{r}x}{2}-\coth\frac{x}{2}\right)\Bigg\}
~,
\end{eqnarray}
in which $\{\frac{\pi}{m_r}\}$ denotes the list of the billiard's corner angles, ${\cal L}$ is the perimeter of the billiard, and
\begin{eqnarray}\label{SelbergBillPo}
\rho_{po}(E)=\Re\left\{\sum_{\{\po\}}\frac{L_{\po}}{4\pi k}\sum_{r=1}^\infty\frac{e^{ikrL_{\po}-ir\gamma_{\po}}}{\sinh \frac{r}{2}u_{\po}}\right\}
~,
\end{eqnarray}
where $\{\po\}$ denotes the list of all primitive periodic orbits, even, odd or singular. In the last equation, the definitions:
\begin{eqnarray}
\label{neg_u_p_maslov}
\begin{array}{cc}
u_{\po}=\left\{\begin{array}{ll} L_p&~~ \mbox{even p.p.o}\\
						L_q+i\pi&~~\mbox{odd p.p.o}\\
						2L_s&~~\mbox{singular p.p.o ,}\\
		      \end{array}\right.
&
\gamma_{\po}=\left\{\begin{array}{ll} 0&~~ \mbox{even p.p.o}\\
						\pi/2&~~\mbox{odd p.p.o}\\
						 L_s/2i&~~\mbox{singular p.p.o .}\\
		      \end{array}\right.\\
\end{array}
\end{eqnarray}
have been used. As for billiards with periodic boundary conditions, one can derive Weyl's asymptotic expansion for the counting function for eigenvalues for hard wall billiards using Selberg's trace formula \Eq{SelbergBill}. For Dirichlet boundary conditions the result is (BV VII.50):
\begin{equation}
\label{weyl_b}
\smoothN(E)\sim \frac{{\cal A}}{4\pi}~E~-~\frac{{\cal L}}{4\pi}\sqrt{E}~-~\frac{{\cal A}}{12\pi}~+~\sum_{\{\frac{\pi}{m_r}\}}\frac{1}{24}\left(m_r-\frac{1}{m_r}\right)+\cdots~~~\mbox{as}~~E\rightarrow\infty
~.
\end{equation}

\subsection{Spectral determinant}\label{sec_spec_det_sel}
Because the Selberg trace formula, that holds for billiards on surfaces of constant negative curvature,  is exact, while Gutzwiller's trace formula is valid only in the semiclassical approximation for hard chaos systems, many of the semiclassical expressions cited in \Sec{SC} are exact for tiling billiards on the pseudo-sphere. For example, \EQ{Delta_prod} is an exact equality for the spectral determinant if one uses the definitions in \Eq{neg_u_p_maslov}. Therefore, \Eq{Delta_sum} is also an exact equality. On the other hand, resumming the problematic sum over primitive periodic orbits in \Eq{Delta_sum} yields only a semiclassical approximation to the exact result. \EQ{Delta_summed} must be adapted to the unit convention and billiards used here. In the units being used $\hbar=1$ so $S=kL$, where $L$ is the length of a trajectory, and the wavenumber $k$ plays the role of $\frac1\hbar$, becoming large in the semiclassical approximation.

Replacements for \EQ{spec_def_a} and \EQ{spec_def_b} are also required:
\bseq
\beqa\label{changes1}
\xi_\mu(\hbar,E)\rightarrow \xi_\mu(k)\equiv L_\mu-\frac{{\cal A}k}{2}~,\\
\label{changes2}
Q^2(K,\frac1\hbar)\rightarrow \frac{Q^2(K,k)}{k}\equiv K^2~+~i\frac{{\cal A}k}{2}~,
\eeqa
where $L_\mu=\sum_{\{j_\po\}}j_\po L_\po$.
\eseq
The result of the resummation for a particular hard wall billiard (defined in \Fig{tile}) is shown in \Fig{fig_spec}, together the exact zeros of the spectral determinant, calculated numerically. One can see that the agreement is excellent even at the lower end of the spectrum.

\setcounter{equation}{0}\def\theequation{\arabic{section}.\arabic{equation}}
\section{Green's function and eigenfunctions on the pseudo-sphere}\label{wave}
In this section the diagonal part of Green's function for billiards on a surface of constant negative curvature is calculated. More precisely, it is shown that for $\zeta=\zeta'$ the general results (\ref{Green_group}, \ref{Green_exact}, \ref{Green_exact2} and \ref{Green_hardwall}) can be written in the form \Eq{ratio0}, namely, as a numerator that depends on the coordinate $\zeta$ over a denominator that is the spectral determinant. Both numerator and denominator are expressed in terms of contributions of various classical orbits. In this way, a comparison to standard semiclassical results is made possible. The eigenfunctions can be calculated directly from the expression of Green's function that is obtained in this section, with the help of \Eq{generalGreen}.

At this point a clarifying remark is due, because Green's functions are ill-defined for ${\bf r}={\bf r'}$ \cite{Morse53}. Green's functions satisfy the equation:
\begin{eqnarray*}
\label{GreensEqn}
(\widehat{\triangle}+E)~G_E({\bf r},{\bf r'})=\delta({\bf r}-{\bf r'})
~,
\end{eqnarray*}
and in two dimensions this necessarily means that the Green function has to have a logarithmic singularity as ${\bf r}\rightarrow{\bf r'}$. This singularity can be removed by subtracting the divergent part:
\begin{eqnarray*}
\label{lim}
G_E({\bf r},{\bf r})\equiv\lim_{\ABS{{\bf a}}\rightarrow{\bf 0}}\left[{G_E({\bf r},{\bf r}+{\bf a})-\frac1{2\pi}\ln{d_{{\bf r},{\bf r}+{\bf a}}}}\right]
~.
\end{eqnarray*}
This subtraction does not affect the spectrum, since it does not depend on $k$.

As in \Sec{sec_density}, it is instructive to start by considering a billiard with periodic boundary conditions on the pseudo-sphere. A generalization for the case of a billiard with Dirichlet boundary conditions is presented in \Sec{appn_wave_bill}. Beginning with the exact (retarded) Green function for $\zeta=\zeta'$, given by \Eq{Green_PBC}:
\begin{eqnarray}\label{green}
G_E(\zeta,\zeta)&=&\frac{-1}{2\pi}Q_{-\half - ik}\left[1\right]~+~\frac{-1}{2\pi}\sum_{p}\sum_{r=1}^{\infty}\sum_{h\in\Gamma/\Gamma_p}Q_{-\half - ik}\left[\cosh {d_{h^{-1}p^rh}(\zeta)}\right]
~.
\end{eqnarray}
Notice that had the above mentioned definition of $G_E({\bf r},{\bf r'})$ not been used, the right hand side of the last equation would include a logarithmic singularity, because (BV G.20):
\beqar
\Re{Q_{-\half-ik}\left[\cosh {d}\right]}\sim-\ln{d}
\as d0~.
\eeqar 

In \Sec{neg} it was noted that the sum over hyperbolic isometries in \Eq{green} contains information about the lengths of the periodic orbits of the billiard. In this chapter more information is going to be pried out of it, in order to sum the repetitions. This will be achieved with the aid of \Eq{beardoneven}. This simple relation from hyperbolic geometry will facilitate the separation of the two different contributions to the distance between a point, $\zeta$, and it's image under a hyperbolic isometry, $g(\zeta)$. One contribution to this distance is the distance of $\zeta$ from the invariant geodesic of $g$, $d_g^\perp(\zeta)$. The second contribution is the length of the periodic orbit that corresponds to the isometry, $L_g$. The treatment of this latter contribution is going to be of vital importance in the ensuing discussion.

\subsection{An exact formula for eigenfunctions}
In order to obtain a transparent formula for Green's function, Legendre functions, which constitute it, are expanded in powers of $e^{-L_g}$ (see \App{Q}):
\begin{eqnarray}\label{legendre-g2}
\hspace{-2cm}Q_{-\half-ik}&&\left[\cosh{d_{g}(\zeta)}\right]=
\sum_{m=0}^\infty{\cal F}_m\left[k,d^\perp_g(\zeta)\right]e^{-\left(\half+m-ik\right)L_g}
~,
\end{eqnarray}
where
\begin{eqnarray}\label{F}
{\cal F}_m\left[k,d^\perp_g(\zeta)\right]\equiv&&
\sum_{n=0}^{\left[\frac{m}{2}\right]}\frac{(2n-1)!!}{(2n)!!}B\left[n+\half-ik,\half\right]  
\times \Lf&&\hspace{-1.5cm}\times
a_{m-2n}\left[\frac{1}{\Cg},\partial \eta\right]\left(\frac{e^{-\eta}}{\Cg}\right)^{\half+2n-ik}  _{\big|_{\eta=0}}
~,
\end{eqnarray}
in which $B(x,y)=\frac{\Gamma(x)\Gamma(y)}{\Gamma(x+y)}$ is the Beta function, and $a_{m}\left[x,y\right]$ is the product two real polynomials of order $m$, one of $x$ and one of $y$.

Returning to \Eq{green} and using \Eq{legendre-g2}, one obtains:
\begin{eqnarray}\label{schematic_green}
G_E(\zeta,\zeta)&=&\frac{-1}{2\pi}Q_{-\half - ik}\left[1\right]~+~
\Lf&&\hspace{1cm}
\frac{-1}{2\pi}\sum_{\{p\}}\sum_{r=1}^{\infty}\sum_{h\in\Gamma /\Gamma_p}\sum_{m=0}^\infty{\cal F}_m\left[k,d^\perp_{h^{-1}p^{r}h}(\zeta)\right]e^{-\left(\half+m-ik\right)L_{h^{-1}p^{r}h}}
~,
\end{eqnarray}
where ${\cal F}_m[x,y]$ was defined in \Eq{F}. 

It was already stated (in the paragraphs before \EQ{Green_exact2}) that all the members of a conjugacy class are related to one periodic orbit. That fact, together with the explanation in the paragraphs preceding \Eq{beardoneven}, leads to:
\beq\label{rL_p}
L_{h^{-1}p^{r}h}=L_{p^{r}}=rL_p
~.
\eeq

Consider the set of distances $\Big\{d^{\perp}_{h^{-1}p^{r}h}(\zeta)\Big\}_{h\in\Gamma/\Gamma_p}$ of the point $\zeta$ from various invariant geodesics. The most obvious point to notice is that,
\beq\label{dperp}
d^{\perp}_{h^{-1}p^{r}h}(\zeta)=d^{\perp}_{h^{-1}ph}(\zeta)
~,
\eeq
because $p^{r}$ and $p$ have the same invariant geodesic.

Another point worth mentioning is that the set of distances $\Big\{d^{\perp}_{h^{-1}ph}(\zeta)\Big\}_{h\in\Gamma/\Gamma_p}$ is, actually, the set of distances of $\zeta$ from all different images of the invariant geodesic of $p$\footnote{The only exception to the rule occurs in hard wall billiards because they have self retracing periodic orbits. Then the list of distances is doubled, because images of $p^{-1}$ are in the conjugacy class.}. This is true because if two boosts ($g_1~\&~g_2$) have the same invariant geodesic they must be repetitions of one another. The reason is the one-to-one correspondence of boosts and periodic orbits: if two periodic orbits overlap, they must be repetitions of one another, because different classical trajectories never cross in phase space. In a billiard with periodic boundary conditions this means that $g_1$ and $g_2$ are identical\footnote{In the Dirichlet case $g_1=g_2^{-1}$ if $g$ corresponds to a self retracing periodic orbit}. 

Now the summation over the distances from the invariant geodesics of members of a hyperbolic conjugacy class may be given a geometrical interpretation. What one actually does in this summation, is to take into account the distance of a point from all of the images of the invariant geodesics of the class representative.

The next aim is to derive \EQ{green_ratio}, \ref{Nsmooth} and \ref{Nosc}, which express the exact Green function in terms of sums over periodic orbit lengths. This will be achieved by using \EQ{rL_p} and \ref{dperp} in \EQ{schematic_green}:
\begin{eqnarray}\label{schematic_green2}
G_E(\zeta,\zeta)&=&
     \frac{-1}{2\pi}Q_{-\half - ik}\left[1\right] +\nonumber \\ 
    &&+\frac{-1}{2\pi}\sum_{\{p\}}\sum_{r=1}^{\infty}\sum_{h\in \Gamma/\Gamma_{p}}\sum_{m=0}^{\infty}{\cal F}_m\left[k,d^\perp_{h^{-1}ph}(\zeta)\right]e^{-\left(\half+m-ik\right)rL_{p}}
~.
\end{eqnarray}
Note that the contribution of a single periodic orbit ($p$) has been decoupled into two separate parts~- the perpendicular distance, $d^\perp_{h^{-1}ph}(\zeta)$, and the length of the periodic orbit, $L_p$. To achieve this in a typical chaotic system one usually has to resort to the semiclassical approximation, while here it is exact. 

Exchanging the order of summation, a geometric series is encountered:
\beq\label{repSum}
{\cal S}^{(m)}(L_p,k)\equiv \sum_{r=1}^{\infty}\left[e^{-(\half+m-ik)L_{p}}\right]^{r}=\frac{e^{-(\half+m-ik)L_{p}}}{1-e^{-(\half+m-ik)L_{p}}}
~.
\eeq
This enables one to write the retarded Green function \Eq{schematic_green2} in the form:
\beq\label{greenLp}
G_E(\zeta,\zeta)=
     \frac{-1}{2\pi}Q_{-\half - ik}\left[1\right]
    +\frac{-1}{2\pi}\sum_{\{p\}}\sum_{h\in \Gamma/\Gamma_{p}}\sum_{m=0}^{\infty}
{\cal F}_m\left[k,d^\perp_{h^{-1}ph}(\zeta)\right] {\cal S}^{(m)}(L_p,k)
~,
\eeq
that is given in terms of the lengths of the primitive periodic orbits of the system. Taking a common denominator and multiplying numerator and denominator by $e^{-i\pi\smoothN (E)}$, one obtains a convenient exact expression for the retarded Green function:

\begin{equation}\label{green_ratio}
G_E(\zeta,\zeta)=\frac{{\cal N}(\zeta,E)}{\Delta(E)}=\frac{\bar{\cal N}(E)+{\cal N}_{po}(\zeta,E)}{\Delta(E)}
~,
\eeq
where
\beq\label{Nsmooth}
\bar{\cal N}(E)\equiv\frac{-1}{2\pi}Q_{-\half-ik}\left(1\right)\Delta(E)
~,
\eeq
in which $\Delta(E)$ is the spectral determinant, as it is given {\em exactly} by \Eq{Delta_prod} using the definitions of \Eq{neg_u_p_maslov} and the replacements defined in \Sec{sec_spec_det_sel} in order to utilize Berry and Keating's results. In addition: 
\beq\label{Nosc}
{\cal N}_{po}(\zeta,E)\equiv \frac{-1}{2\pi}\sum_{\{p\}}\sum_{h\in \Gamma/\Gamma_{p}}\Sum_{m=0}^{\infty}
{\cal F}_m\left[k,d^\perp_{h^{-1}ph}(\zeta)\right] \Delta^{(p,m)}(E)
~.
\end{equation}
$\Delta^{(p,m)}(E)$ is defined in \Eq{Deltapm_def}. According to \EQ{Delta_sum} and \ref{Deltapm_sum}, as explained in \Sec{sec_spec_det_sel},  $\Delta(E)$ and $\Delta^{(p,m)}(E)$ are given exactly by sums over composite orbits:
\begin{eqnarray*}
\Delta(E)&=&\sum_{\mu}C_{\mu}e^{-i\pi\smoothN (E)+ikL_\mu}~,\\
\Delta^{(p,m)}(E)&=&\sum_{\mu}C_{\mu}^{(p,m)}e^{-i\pi\smoothN (E)+ikL_{\mu,p}}
~.
\end{eqnarray*}
$C_{\mu}$ and $C_{\mu}^{(p,m)}$ are given by their definitions \EQ{Cmu_berry} and \ref{Cmupm}, with \ref{neg_u_p_maslov}. It turns out that all of these coefficients are real numbers (see \App{C}). The $L_\mu$'s were defined after \Eq{changes1}, and $L_{\mu,p}=L_\mu+L_p$.

The criterion for the separation of the numerator in \EQ{green_ratio} into two parts, is that tracing the first term gives the smooth term in the density of states to \Eq{Selberg}, and tracing the second one gives the periodic orbit contribution to \Eq{Selberg}.

There are three sorts of contributions to \EQ{green_ratio}. The first is from short non-periodic orbits, and it is given by $\bar{\cal N}(E)$, as it results from the identity. The second and third contributions are related to periodic orbits and are given by ${\cal N}_{po}(\zeta,E)$. The first contribution to this function is related to the actual periodic orbits in the billiard, while the second is related to their images. The reason is that a ``real'' periodic orbit is made up from only a few segments of invariant geodesics of a conjugacy class's elements. The invariant geodesics of the rest of the conjugacy class make up ``images'' of the periodic orbit, as they never intersect with the original billiard. ``Real'' periodic orbit segments generally have relatively small $d^\perp_{h^{-1}ph}(\zeta)$, while the images of them tend to have larger $d^\perp_{h^{-1}ph}(\zeta)$.

\EQ{green_ratio} is a manifestation of the results of Fredholm's theory, as presented in \cite{Fishman96}. Unlike in the cases considered there, the exact Green function is written as the ratio between two sums over periodic orbit lengths without the need to resort to a semiclassical approximation. The precise nature of the connection with Fredholm's theory has not been worked out yet for this case. Once this issue is resolved, Fredholm's theory will supply a prescription as to the correct reordering of terms in the series, so that they converge absolutely to the exact result. 

The eigenfunctions densities can be found from \Eq{green_ratio} with the help of \Eq{generalGreen}. The poles of Green's function are the eigenvalues of the Hamiltonian, and the exact eigenfunction densities are the residues:
\beq\label{exact_eig}
\left| \psi_{\alpha}(\zeta)\right|^{2}=\Res{\frac{{\cal N}\big(\zeta,E_\alpha\big)}{\Delta\big(E_\alpha\big)}}=\frac{{\cal N}\big(\zeta,E_\alpha\big)}{\Delta'\big(E_\alpha\big)}
~,
\eeq
where $\Delta'\big(E_\alpha\big)$ denotes a derivative with respect to energy of the spectral determinant, evaluated at $E_\alpha$.

In order to make sense of the infinite sums over the periodic orbit lengths resummation is required, as they do not converge absolutely. This is performed in the \Sec{discuss}.

\subsection{Hard wall billiards}\label{appn_wave_bill}
A formula for Green's function for tiling billiards with Dirichlet boundary conditions on the pseudo-sphere that is analogous to \EQ{green_ratio}~-~\ref{Nosc} will be developed in this subsection. The differences from the periodic boundary conditions' case arise from the fact that now there are several new kinds of transformations: inversions ($\cal I$), rotations ($\cal R$) and odd boosts. In principle, the derivation of a formula for Green's function is exactly the same as the previous derivation.  As before, the exact Green function can be written as the ratio between two functions that involve summation over all the periodic orbits of the system. The differences are of technical nature only. 

As before, one may separate Green's function into part that is related to the periodic orbits and to a part that is not (see \EQ{Green_exact2}). The criterion for the separation is the same as mentioned after \Eq{Nosc}, namely tracing the first term gives the smooth term in the density of states \Eq{SelbergBillSmooth}, and tracing the second one gives the periodic orbit contribution \Eq{SelbergBillPo}. According to \Eq{Green_hardwall}, for the Dirichlet boundary conditions case, the latter is given by:
\begin{eqnarray}
G^{po}_E\left(\zeta,\zeta\right) &=&\frac{-1}{2\pi}\Bigg\{
\sum_{\{p\}}\sum_{r=1}^{\infty}\sum_{h\in \Gamma/\Gamma_p}Q_{-\half - ik}\left[\cosh{d_{hp^rh^{-1}}(\zeta)}\right]
+\nonumber\\&&+
\sum_{\{q\}}\sum_{r=1}^{\infty}\sum_{h\in \Gamma/\Gamma_q}(-1)^r~Q_{-\half - ik}\left[\cosh{d_{hq^rh^{-1}}(\zeta)}\right]
+\\&&+
\sum_{\{s\}}\sum_{r=1}^{\infty}\sum_{h\in \Gamma/\Gamma_s}\bigg[Q_{-\half - ik}\left[\cosh{d_{hs^rh^{-1}}(\zeta)}\right]-Q_{-\half - ik}\left[\cosh{d_{h{\cal I}s_{\cal I}^rh^{-1}}(\zeta)}\right]\bigg]\Bigg\}\nonumber~.
\end{eqnarray}
The three types of boosts that exist have been separated:
\begin{enumerate}
	\item $\sum_{\{p\}}\sum_{r=1}^\infty$ is a sum over the classes conjugate to powers of primitive even boosts.
	\item $\sum_{\{q\}}\sum_{r=1}^\infty$ is a sum over the classes conjugate to powers of primitive odd boosts. The factor $(-1)^r$ is necessary, because if $r$ is odd/even $hq^rh^{-1}$ involves an odd/even number of inversions.
	\item $\sum_{\{s\}}\sum_{r=1}^\infty$ is a sum over the classes conjugate to powers of primitive singular boosts. These boosts are the boosts whose invariant geodesic lies along an inversion line of the group, which are images of the boundary of the billiard (see BV Appendix M). This means that $\left[s,{\cal I}_s\right]=\left[{\cal I}_s s,{\cal I}_s\right]=0$. This is why both $s$ and ${\cal I}_s s$ must be included, and why the contributions have opposite signs.
 \end{enumerate}
For the primitive even boosts a contribution identical to \Eq{legendre-g2} is obtained, while for the primitive odd boosts the result is (see \App{Q}): 
\begin{eqnarray}
\label{legendre-q}
\hspace{-0.3cm}(-1)^rQ_{-\half - ik}\left[\cosh{d_{hq^rh^{-1}}(\zeta)}\right]&=&
\Lf&&\hspace{-4cm}=
(-1)^r~
\sum_{m=0}^{\infty} \big((-1)^r\big)^{m} {\cal F}_m\left[k,d^\perp_{h^{-1}qh}(\zeta)\right]e^{-\left(\half+m-ik\right)rL_{q}}
~,
\end{eqnarray}
where ${\cal F}_m\left[k,d^\perp_{h^{-1}{\po}h}(\zeta)\right]$ was defined in \EQ{legendre-g2}. For  the primitive singular boosts a slightly different expression is obtained (see \App{Q}): 
\begin{eqnarray}\label{legendre-s}
Q_{-\half - ik}\big[\cosh{d_{hs^rh^{-1}}(\zeta)}\big]-
Q_{-\half - ik}\big[\cosh{d_{h{\cal I}s^rh^{-1}}(\zeta)}\big]&=&
\Lf&&\hspace{-6cm}=
\sum_{m=0}^{\infty}\left[1-(-1)^{m}\right] {\cal F}_m\left[k,d^\perp_{h^{-1}sh}(\zeta)\right]e^{-\left(\half+m-ik\right)rL_{s}}
~.
\end{eqnarray}
The important fact is that, as in \Eq{legendre-g2}, the contribution of the periodic orbits to the exact Green function is now expressed as a sum over periodic orbits in a form that admits summation over repetitions. Since \EQ{legendre-g2}, \ref{legendre-q} and \ref{legendre-s} have a similar form, the same steps that were undertaken in the previous case apply here. After summing over the repetitions and taking a common denominator, the expression \Eq{green_ratio} is found, but $\bar{\cal N}(E)$ and ${\cal N}_{po}(\zeta,E)$ are of a different form. The function $\bar{\cal N}(E)$ will be treated later, while:
\beq\label{Nosc_bill}
{\cal N}_{po}(\zeta,E)\equiv \frac{-1}{2\pi}\sum_{\{\po\}}\sum_{h\in \Gamma/\Gamma_{\po}}\Sum_{m=0}^{\infty}
\Phi^{\po}_m\left[k,d^\perp_{h^{-1}{\po}h}(\zeta)\right] \Delta^{(\po,m)}(E)
~,
\eeq
where $\sum_{\{\po\}}$ denotes summation over all primitive periodic orbits,  even, odd or singular, and:
\beqar
\Phi^{\po}_m\left[k,d^\perp_{h^{-1}{\po}h}(\zeta)\right] \equiv
\left\{\begin{array}{cl} {\cal F}_m\left[k,d^\perp_{h^{-1}{\po}h}(\zeta)\right] &~~\po\neq s\\
			     2{\cal F}_{2m+1}\left[k,d^\perp_{h^{-1}{\po}h}(\zeta)\right] &~~\po= s\\
		      \end{array}\right.
\eeqar
Using the definitions of  \Eq{neg_u_p_maslov}, $\Delta(E)$ is given by \Eq{Delta_sum} and $\Delta^{(\po,m)}(E)$ by \Eq{Deltapm_sum}.

One can see that the additions to \Eq{Nosc} required for the calculation of the contribution of the primitive periodic orbits to the eigenfunction  density in billiards with hard walls, do not change it's nature, adding only contributions of other types of periodic orbits, namely odd and singular ones, and their corresponding amplitudes and phases.

The contribution of non-periodic orbits is given by the contribution of the non-boost isometries of the group, whose fundamental domain is the billiard under study. In the case of billiards with periodic boundary conditions there was only one such isometry, namely the identity, that corresponds to zero length orbits. In the case of billiards with Dirichlet boundary conditions there are many such isometries besides the identity,  all of them  rotations and inversions, and they correspond to orbits closed in real space, not in phase space. According to \EQ{Green_hardwall}, the contribution of these isometries to Green's function is:
\begin{eqnarray}
\label{greenbilliardsmooth}
\bar{G}_E(\zeta,\zeta)&=&\frac{-1}{2\pi}\bigg\{Q_{-\half - ik}\left[1\right]~
-\sum_{\stackrel{Inversions}{\cal I}} Q_{-\half - ik}\left[\cosh{\di}\right]+ 
\Lf&&\hspace{4cm}+
\sum_{\stackrel{Rotations}{\cal R}} Q_{-\half - ik}\left[\cosh{\dr}\right]\bigg\}
~.
\end{eqnarray}

This means that ${\bar{\cal N}}\left(E\right)$ of \Eq{Nsmooth} should be replaced by:
\begin{eqnarray}
\label{Nbilliardsmooth0}
{\bar{\cal N}}\left(E,\zeta\right)&=&\frac{-1}{2\pi}
\left[Q_{-\half-ik}\left[1\right]~-\sum_{\stackrel{Inversions}{\cal I}} Q_{-\half - ik}\left[\cosh{\di}\right]~+\right.
\Lf&&\hspace{2cm}
\left.+~\sum_{\stackrel{Rotations}{\cal R}} Q_{-\half - ik}\left[\cosh{\dr}\right]\right]\times\Delta(E)
~.
\end{eqnarray}

The main difference between \Eq{Nsmooth} and this result, is that the former is a constant function of position, whereas the latter is not. 

\section{Resummation}\label{sec_resum}
As it stands, \EQ{green} and all equations derived from it are not very useful because of the same convergence difficulties \Eq{Gutzwiller} presented. The reason is the exponential proliferation of the number of images of $\zeta$ with the distance $d_g{(\zeta)}$. When large enough, the distance $d=d_g{(\zeta)}$ is approximately the distance of the the $g$'th tile from the billiard. The number of tiles at such a distance is roughly proportional to the circumference of a circle of radius $d$ around the original billiard. On the pseudo-sphere this circumference is proportional to $e^d$ as ${d}\rightarrow{\infty}$. On the other hand, the behavior of $Q_{-\half-ik}[\cosh d]$ for large $d$ is:
\beq
Q_{-\half-ik}[\cosh d] \sim e^{\left(ik-\half\right) d}\as{d}{\infty}
~,
\eeq
so the net contribution of images at a distance $d$ to \Eq{green} is proportional to
\beq\label{term_growth}
e^{\left(ik+\half\right) d} \as{d}\infty
~.
\eeq
This means that the sum in \Eq{green} is at best only conditionally convergent, which is why resummation is needed to make use of it.  

Notice that most of the exponential proliferation that causes the convergence problems comes from the exponential proliferation of classes conjugate to primitive isometries, which is equivalent to the exponential proliferation of long primitive periodic orbits. This is because the number of periodic orbits of length $d$ is \cite{Huber59,Balazs86}:
\beq\label{exp_prolif}
N(d)\sim \frac{e^d}{d} \as{d}{\infty}
~,
\eeq
which means that the number of repetitions of short primitive periodic orbits is sub-dominant and that the number of terms in the sum over a conjugacy class's elements grows sub-exponentially. Resummation following Berry and Keating \cite{Berry92} will be applied to get a meaningful result from this sum. It will result in an approximation on the level of the semiclassical approximation

Due to \EQ{term_growth}, the sum over $\{p\}$ in \EQ{green_ratio} does not converge absolutely for real $k$'s. On the other hand, if: 
\beq\label{entropy_barrier}
\Im k>\half
\eeq
it does. It will be assumed that ${\cal N}\big(\zeta,E(k)\big)$ is analytic in a sufficiently wide strip around the $\Re k$ axis to resum the problematic sum by analytic continuation from the region described by \Eq{entropy_barrier} to the region of physical interest $\Im k=0$.

In the previous section, a formula in terms of periodic orbits was obtained for the exact expression:
\begin{equation}\label{wavesum}
G_E(\zeta,\zeta)=\sum_{\alpha} \frac{\left| \psi_{\alpha}(\zeta)\right|^{2}}{E(k)-E_\alpha}=
\frac{{\cal N}\big(\zeta,E(k)\big)}{\Delta\big(E(k)\big)}
~.
\end{equation}
Resummation of the denominator was discussed in \Sec{sec_spec_det_sel}, so only the numerator will be treated explicitly in this section.

Since $E(k)=k^2+\qrt=E(-k)$, and since \cite{Berry92}:
\beq
\Delta{\big(E(k)\big)}=\Delta{\big(E(-k)\big)}
~,
\eeq
then:
\beq
{\cal N}\big(\zeta,E{(k)}\big)={\cal N}\big(\zeta,E({-k})\big)
~.
\eeq
This exact symmetry is crucial for the analytic continuation method of resummation to work.

Using Cauchy's theorem one may write:
\beq
\label{analy}
{\cal N}\big(\zeta,E{(k)}\big)=\frac{1}{2\pi i} \oint_{C} \frac{dz}{z} \gamma (z,k) {\cal N}\big(\zeta,E(k+z) \big)
~.
\eeq
Leading to:
\begin{eqnarray}
\label{NandU}
{\cal N}\big(\zeta,E{(k)}\big) &=& \frac{1}{2\pi i}\Int_{\infty+i(1/2+\varepsilon)}^{-\infty+i(1/2+\varepsilon)}
\frac{dz}{z} \gamma (z,k) \left[{\cal N}\big(\zeta,E{(z+k)} \big)+{\cal N}\big(\zeta,E{(z-k)} \big)\right]
~,
\end{eqnarray}
where the {\it ad hoc} choice of Berry and Keating \cite{Berry92} will be used:
\begin{equation}
\label{gamma}
\gamma (z,k)=e^{-\frac{1}{2k} K^2 z^2}
~.
\end{equation}

As shown in \App{resum}, \EQ{NandU}'s leading order approximation is given by:
\bseq
\beq
{\cal N}\big(\zeta,E{(k)}\big)=\bar{\cal N}\big(E{(k)}\big)+{\cal N}_{po}\big(\zeta,E{(k)}\big)
\as k\infty~,
\eeq
where
\beq
\label{Nsmooth_final}
\bar{\cal N}\big(E{(k)}\big)\sim\frac{-1}{2\pi} \Re\Bigg\{
     Q_{-\half-ik}[1]
\left({\bar{\Delta}(E)+i\bar{\Delta_i}(E)}\right)\Bigg\}
\as k\infty~.
\eeq
$\bar{\Delta}(E),~\bar{\Delta_i}(E)\in\real$ are given by:
\beq
{\bar{\Delta}(E)+i\bar{\Delta_i}(E)}\equiv{\sum_\mu C_\mu e^{-i\pi\smoothN (E)+ikL_\mu}
\Erfc\left[\sqrt{\frac{k}{2Q^2(K,k)}}\xi_\mu(k)\right]}
~,
\eeq
\eseq
where $\xi_\mu(k)$ and $Q^2(K,k)$ are given by \Eq{changes1} and \Eq{changes2}.

In addition:
\bseq
\begin{eqnarray}
\label{Nosc_final}
{\cal N}_{po}&&\big(\zeta,E{(k)}\big)\sim\frac{1}{2\pi} \Im\Bigg\{\sum_{\{p\}}\sum_{h\in \Gamma/\Gamma_{p}}\bar{\Delta}^{(p,h)}(\zeta,k)\Bigg\}
\as k\infty~,
\end{eqnarray}
where:
\beqa\label{delta_ph}
\bar{\Delta}^{(p,h)}(\zeta,k)&=&\sum_\mu \Delta^{(p,h)}_{\mu}(\zeta,k) e^{-i\pi\smoothN (E)+ikL_{\mu,p}}\Erfc\left[\sqrt{\frac{k}{2Q^2(K,k)}}\xi_{\mu,p}^{h}(k,\zeta)\right]
~,
\eeqa
in which $\xi_{\mu,p}^{h}(k,\zeta)\equiv L_{\mu,p}+\ln\Chp-\half k_\alpha{\cal A}$, and:
\beqa\label{Delta_phmu_approx}
\Delta^{(p,h)}_{\mu}(\zeta,k)&\sim&\sqrt{\frac{\pi}{2ik}}\sum_{j=0}^{j_{p}}\frac{\bar{C}_\mu^{(p,j)}e^{ik\left( d_{hp^{j+1}h^{-1}}\left(\zeta\right)-(j+1)L_p\right)} }{\sqrt{\sinh{d_{hp^{j+1}h^{-1}}\left(\zeta\right)}}}
\as{k}{\infty}
~,
\eeqa
where $\bar{C}_\mu^{(p,j)}$ was defined in \Eq{def_Cmupj}, and the definitions of \Eq{neg_u_p_maslov} for even boosts were used.
\eseq

Since  both sides of \Eq{wavesum} have only simple poles at the eigenvalues of the Hamiltonian, the residue at $E=E_\alpha$ is $\left| \psi_{\alpha}(\zeta)\right|^{2}$. Hence one obtains:
\beq
\label{psi}
\left| \psi_{\alpha}(\zeta)\right|^{2}\sim \overline{\left| \psi_{\alpha}(\zeta)\right|}^{2}+\left| \psi_{\alpha}(\zeta)\right|_{po}^{2}
\as{k}{\infty}~,
\eeq
where
\beqa
\label{psiSmooth}
\overline{\left| \psi_{\alpha}(\zeta)\right|}^{2}&\equiv&\frac{\bar{\cal N}(\zeta,E_{\alpha})}{\Delta'(E_\alpha)} ~,\\
\label{psiOsc}
\left| \psi_{\alpha}(\zeta)\right|_{po}^{2}&\equiv&\frac{ {\cal N}_{po}(\zeta,E_{\alpha})}{\Delta'(E_\alpha)}
~.
\eeqa
Notice that on the spectrum:
\begin{eqnarray}
\label{Nsmooth_spec}
\bar{\cal N}&&(\zeta,E_\alpha)=\qrt{\bar{\Delta}_i(E_\alpha) \tanh{\pi k_\alpha}}
~.
\end{eqnarray}
In \Eq{delta_ph} an effective truncation of the problematic sums over periodic orbits ($\equiv$conjugacy classes) in \Eq{Nosc} was obtained. The cut-off is centered around
\beq
\xi_{\mu,p}^{h}(k_\alpha,\zeta)=0 
~,
\eeq
that corresponds to 
\beq
L_{\mu,p}+\ln\Chp=\frac{k_\alpha{\cal A}}2
~,
\eeq
which is equal to the distance traversed during half the Heisenberg time. There are two sorts of contributions to the sum over periodic orbits. The first is from geodesics that pass through the billiard. Segments of these  geodesics constitute the ``real'' periodic orbits of the system, so they are characterized by relatively small $d_{hph^{-1}}^\perp$'s. The second contribution is from geodesics that never intersect the billiard. They only pass through copies of it, so they are characterized by relatively large values of $d_{hph^{-1}}^\perp$'s. This is why the first contribution can be associated with periodic orbits, and the second with their images. These two sorts of contributions are translated to two sorts of cut-offs. The first is the cut-off of long periodic orbits, which is encountered in typical chaotic systems too, while the second is the cut-off of distant images of short periodic orbits.

\subsection{Resummation for hard wall billiards}\label{sec_resum_bill}
As for billiards with periodic boundary conditions, in order for \Eq{green_ratio} to be meaningful for hard wall billiards, resummation is required, because \EQ{Nosc_bill} and \ref{Nbilliardsmooth0} involve sums that do not converge absolutely. The same scheme Berry and Keating \cite{Berry92} introduced may be used again. The only qualitative difference results from Weyl's asymptotic expansion for the counting function of eigenenergies \Eq{weyl_b} because it's high order coefficients are not identically zero, as they were in \Eq{weyl_a}. Despite of this, in the leading order approximation, there is no difference. The result of the resummation is, to leading order:
\begin{eqnarray}
\label{Nosc_final_bill}
{\cal N}_{po}&&\big(\zeta,E{(k)}\big)\sim\frac{1}{2\pi} \Im\Bigg\{\sum_{\{\po\}}\sum_{h\in \Gamma/\Gamma_{\po}}\bar{\Delta}^{(\po,h)}(\zeta,k)\Bigg\}\as{k}\infty
~,
\end{eqnarray}
where:
\beqa\label{delta_ph_bill}
\bar{\Delta}^{(\po,h)}(\zeta,k)&=&\sum_\mu \Delta^{(\po,h)}_{\mu}(\zeta,k) e^{-i\pi\smoothN (E)+ikL_{\mu,\po}}\Erfc\left[\sqrt{\frac{k}{2Q^2(K,k)}}\xi_{\mu,\po}^{h}(k,\zeta)\right]
~.
\eeqa
\EQ{delta_ph_bill} includes the contribution of \EQ{Delta_phmu_approx} (even boosts), odd boosts:
\bseq
\beqa\label{Delta_qhmu_approx}
\Delta^{(q,h)}_{\mu}(\zeta,k)&\sim&\sqrt{\frac{\pi}{2ik}}\sum_{j=0}^{j_{q}}(-1)^{j+1}\bar{C}_\mu^{(q,j)}e^{i(j+1)\gamma_q}\frac{e^{ik_{\alpha}\left( d_{hq^{j+1}h^{-1}}\left(\zeta\right)-(j+1)L_q\right)} }{\sqrt{\sinh{d_{hq^{j+1}h^{-1}}\left(\zeta\right)}}}
\as{k}\infty,
\eeqa
as well as singular boosts:
\beqa\label{Delta_shmu_approx}
\Delta^{(s,h)}_{\mu}(\zeta,k)&\sim&\sqrt{\frac{\pi}{2ik}}\sum_{j=0}^{j_{s}}\bar{C}_\mu^{(s,j)}e^{i(j+1)\gamma_s}e^{-ik(j+1)L_s}
\times\Lf&&\hspace{-1cm}\times
\bigg[\frac{e^{ik_{\alpha}d_{hs^{j+1}h^{-1}}\left( \zeta\right)}} {\sqrt{\sinh{d_{hs^{j+1}h^{-1}}\left(\zeta\right)}}}-\frac{e^{ik_{\alpha} d_{h{\cal I}_ss^{j+1}h^{-1}}\left(\zeta\right)}} {\sqrt{\sinh{d_{h{\cal I}_ss^{j+1}h^{-1}}\left(\zeta\right)}}}\bigg]
\as{k}{\infty}
,
\eeqa
\eseq
where the coefficients  $\bar{C}_\mu^{(q,j)}$ and $\bar{C}_\mu^{(s,j)}$ are those of \Eq{def_Cmupj} with the definitions of \Eq{neg_u_p_maslov}.

As to the contribution of the non-periodic orbits, \Eq{Nbilliardsmooth0} leads, after the analytic continuation, to the expression that
should replace \Eq{Nsmooth_spec} :
\begin{eqnarray}
\label{Nbilliardsmooth}
{\bar{\cal N}}\left(E_\alpha,\zeta\right)&=&\frac{\bar{\Delta_i}(E_\alpha) \tanh{\pi k_\alpha}}{4}+
\\&&
+\frac{1}{2\pi}\Re\Bigg\{\sum_{\mu}C_{\mu}e^{-i\pi\smoothN (E_\alpha)+i k_\alpha L_\mu}\times
\nonumber\\&&
\times\bigg[-\sum_{\stackrel{Inversions}{\cal I}} Q_{-\half - ik_\alpha}\left[\cosh{\di}\right]\Erfc\Big[\sqrt{\frac{k_\alpha}{2Q^2(K,k_\alpha)}}\left(\xi_\mu(k_\alpha)+\di\right)\Big]+
\Lf&&
+\sum_{\stackrel{Rotations}{\cal R}} Q_{-\half - ik_\alpha}\left[\cosh{\dr}\right]\Erfc\Big[\sqrt{\frac{k_\alpha}{2Q^2(K,k_\alpha)}}\left(\xi_\mu(k_\alpha)+\dr\right)\Big]
\bigg]\Bigg\}
\nonumber~.
\end{eqnarray}

\section{Numerical calculation of eigenfunctions}\label{Sec_numerics}
\EQ{green_ratio} for billiards with periodic boundary conditions (with \ref{Nsmooth} and \ref{Nosc}), or for hard wall billiards (with \ref{Nosc_bill} and \ref{Nbilliardsmooth0}), is exact. It cannot be used, however, for the calculation of eigenfunctions, since both the numerator and the denominator are expressed as series that are not absolutely convergent. In what follows, resummed expressions obtained from the above equations, namely \EQ{psi} with \ref{Nosc_final_bill} and \ref{Nbilliardsmooth} will be calculated and compared with exact eigenfunctions found numerically. In particular, an arbitrary slice through the two lowest eigenfunctions of the hard wall billiard defined in \Fig{tile} is calculated. This billiard tiles the pseudo-sphere perfectly by reflections, because all of it's angles are of integer fractions of $\pi$ \cite{Coxeter65}. At such low energies, the semiclassical approximations used in the resummation are not expected to be of very high accuracy. The reason higher eigenfunctions were not investigated is that a large number of isometries has to be used in the calculation. There are billiards for which it is very easy to obtain large numbers of isometries. These billiards are called ``Arithmetic Billiards'', and the reason for not choosing one of them is that some of their physical properties, such as the energy spectrum and the spectrum of classical periodic orbit lengths, is non-generic \cite{Bogomolny97}.

The results are shown in \Fig{wavecut1} and \ref{wavecut2}. Presented are the exact cut through the eigenfunction density, the contribution of the the non-boost isometries of \Eq{Nbilliardsmooth}, and the corrections to this due to the periodic orbits of \Eq{Nosc_final_bill}. The contribution of the non-boost isometries accounts for most of the features of the eigenfunction density, while the addition of the periodic orbits provides only a small correction. 

One can see that even after the inclusion of all of the isometries the sums truncated by resummation require, the eigenfunctions are still not reproduced perfectly. A measure of the quality of an approximation of the function $f(x)$ by $f_{approx}(x)$ is given by the average squares deviation:
\beq\label{diff}
\AV{D}\equiv\frac{\int~dx \ABS{f(x)-f_{approx}(x)}^2}{\int~dx \ABS{f(x)}^2}
~,
\eeq
where in this case $f(x)=\ABS{\Psi(\zeta)}$, and the integral is taken along the cut (see \Fig{wavecut1}).  It is interesting to see the difference in the quality of the approximation without including the periodic orbit contribution, $\AV{D_1}$, and including it, $\AV{D_2}$. For the ground-state (see \Fig{wavecut1}):
\beqar
\AV{D_1}&\approx&0.023~,\\
\AV{D_2}&\approx&0.015~.
\eeqar
For the first excited state the result is:
\beqar
\AV{D_1}&\approx&0.043\\
\AV{D_2}&\approx&0.030.
\eeqar
These numbers show that the quality is fairly good even without including the periodic orbits' contribution. Including this last contribution improves the approximation appreciably. The reason the fit is not perfect is that at such low energies the number of periodic orbits that contribute to the truncated sums is very small. The cut-off region extends all the way from $L_\po=0$ to $L_\po=2.00$ for the ground-state and to $L_\po=2.46$ for the first excited state. This means that almost all the periodic orbits that contribute, in both cases, to the eigenfunction density are from the cut-off region. Therefore, one cannot expect that the truncated sums in \Eq{delta_ph_bill} and \Eq{Nbilliardsmooth} should do too well, because in deriving them, terms of the same order were neglected, namely terms supported by the cut-off region exclusively. For such low levels it is essential to include high order corrections to the truncated sums, and although it is possible to do so in an obvious manner, it was not done to date.

\setcounter{equation}{0}\def\theequation{\arabic{section}.\arabic{equation}}
\section{A comparison with previous results}\label{discuss}
One observes that the structure of the resummed formula for the periodic orbit contribution to eigenfunctions obtained in the present work \Eq{psiOsc}, and the analogous expression \Eq{psi-O} Agam and Fishman \cite{Agam93} obtained is very similar. Therefore, it is illuminating to compare the two, because the methods of derivation differ qualitatively. 

First, the expression  \Eq{psi-O} Agam and Fishman obtained  in terms of the quantities that apply for tiling billiards with periodic boundary conditions on the pseudo-sphere will be presented. For this purpose, the quantities identified in \EQ{changes1}, \ref{changes2} and the following, are required:
\begin{enumerate}
	\item $\dot q_{||}\sim2k\as k\infty$;
	\item From \EQ{Rp} and the monodromy matrix for billiards with periodic boundary conditions (see Eq.~42 in \cite{Aurich91}): $\half\left[R^+_\po-R^-_\po\right]\sim k\as k\infty$;
	\item $u_\po=L_p$ and $\gamma_\po=0$, because the simple case of periodic boundary conditions is being considered (see \EQ{neg_u_p_maslov});
	\item From \EQ{xi_mup_def} : $\xi_{\mu,\po}(\hbar,E)=kL_{\mu,\po}-\frac{{\cal A}k^2}{2}$.
\end{enumerate}
Substituting these in \Eq{psi-O}, one obtains:
\beqa\label{psi-O-bill}
\ABS{\Psi_\alpha({\bf q})}^2_{po} &\sim& \frac{1}{2\pi k_{\alpha}\Delta^{\prime}(E_\alpha)}\Im\Bigg\{\sqrt{\frac{\pi k_{\alpha}}{2i}}\sum_{\{p\}}\sum_\mu \sum_{j=0}^{j_p}
\frac{e^{  ik_{\alpha} q_{\perp}^2\tanh{\frac{j+1}2L_{p}}}}{\sqrt{\sinh{(j+1)L_p}}}
\bar{C}_\mu^{(p,j)}
 \times\\
&&\!\!\!\!\!\!\!\!\!\!\!\!\!\!\!\times e^{-i\pi\smoothN(E_{\alpha})+ik_{\alpha}L_{\mu,p}}\Erfc\left[ \sqrt{\frac{k_{\alpha}}{{2 Q^2(K,k_{\alpha})}}}\left({L}_{\mu,p}  + q_{\perp}^2-\frac{{\cal A}k_\alpha}{2}\right)  \right]\Bigg\}
\as{k_\alpha}{\infty}~,\nonumber
\eeqa
where $\{p\}$ denotes the list of primitive periodic orbits of the billiard. We turn now to write the result \Eq{psiOsc}, obtained in this work, in a form that can be compared with \Eq{psi-O-bill}. The asymptotic expansion that led to \Eq{psi-O} was obtained after it was argued that the leading order contribution was from a small region around periodic orbits \cite{Berry89a}. Therefore, the relevant part of \Eq{psiOsc} in such a region should be found. It is easy to see that what remains of $\sum_{\{p\}}\sum_{h\in \Gamma/\Gamma_{p}}$ is the sum over the isometries whose invariant geodesics constitute the actual trajectory in the billiard, because images of periodic orbits lie outside of the billiard, and therefore there are no points in their close vicinity. The remaining sum will be denoted by $\sum\limits_{p.o.}$. The next stage is to obtain an approximate expression for the $\zeta$-dependent part of \Eq{psiOsc}. For $d_g^{\perp}(\zeta)\rightarrow0$, it turns out that (from \EQ{beardoneven}):
\beq
d_g(\zeta) \sim L_g+\tanh{\half L_g}~{d_g^{\perp}}^2(\zeta)
\as {d_g^{\perp}(\zeta)}0
~.
\eeq
This means that, to leading order:
\bseq
\beq\label{Qapprox}
\frac{e^{ik_{\alpha}\left( d_{hp^{j+1}h^{-1}}\left(\zeta\right)-(j+1)L_p\right)}}{\sqrt{\sinh{d_{hp^{j+1}h^{-1}}\left(\zeta\right)}}} 
\sim
\frac{ e^{ik_{\alpha}\tanh{\half (j+1) L_p}~{d_{p.o.}^{\perp}}\hspace{-0.2cm}{}^2(\zeta)}}{\sqrt{\sinh{(j+1)L_{p}}}}
~,
\eeq
and
\beq\label{dperpapprox}
\ln{\cosh{{d_{p.o.}^{\perp}}\hspace{-0.2cm}{}^2(\zeta)}} \sim {d_{p.o.}^{\perp}}\hspace{-0.2cm}{}^2(\zeta)
\as {d_{p.o.}^{\perp}(\zeta)}0~.
\eeq
\eseq
Inserting these into \Eq{psiOsc}, one obtains:
\beqa\label{psi_d_small}
\ABS{ \psi_{\alpha}(\zeta)}^{2}_{po} &\sim& \frac{1}{2\pi k_{\alpha}\Delta'(E_\alpha)}  \Im\Bigg\{ \sqrt{\frac{\pi k_{\alpha}}{2i}}
\sum_{p.o.}\sum_{\mu}  \sum_{j=0}^{j_p} \frac{ e^{ik_{\alpha}\tanh{\half (j+1) L_p}~{d_{p.o.}^{\perp}}\hspace{-0.2cm}{}^2(\zeta)}}{\sqrt{\sinh{(j+1)L_{p}}}}  \bar{C}_\mu^{(p,j)}
\times\\&&\times
e^{-i\pi\smoothN (E)+ik_{\alpha}L_{\mu,p}}\Erfc\left[\sqrt{\frac{k_{\alpha}}{2Q^2(K,k_\alpha)}}\left(L_{\mu,p}+{d_{p.o.}^{\perp}}\hspace{-0.2cm}{}^2(\zeta)-\frac{{\cal A}k_\alpha}{2}\right)\right] \Bigg\}\nonumber\\ &&
\;\;\;\;\;\;\;\;\;\;\;\;\;\;\;\;\;\;\;\;\;\;\;\;\;\;\;\;\;\;\;\;\;\;\;\;\;\;\;\;\;\;\;\;\;\;\;\;\;\;\;\;\;\;
\as{k_\alpha}{\infty}\mbox{ and }{d_{p.o.}^{\perp}(\zeta)}\rightarrow0~.\nonumber
\eeqa
One can see that \Eq{psi-O-bill} and \Eq{psi_d_small} are exactly the same, although the derivation of these two relations was completely different.

The above comparison brings to light what is missing in the result of Agam and Fishman \cite{Agam93}. Their analysis was based on an expansion around periodic orbits, while in the derivation of \Eq{psiOsc} here, there was no such assumption. In particular, \Eq{psiOsc} includes the contribution of {\em images} of periodic orbits, which one would never expect to obtain from an expansion around periodic orbits, because these images tend to lie far away from points in the billiard.

The expression \Eq{psi_d_small} obtained near periodic orbits reveals the role of periodic orbits in the eigenfunction density. As the distance from a periodic orbit is increased, fringes occur, that oscillate on a scale of $\frac{1}{\sqrt{k}}$, and as $k\rightarrow\infty$ they are washed out, as a result of contributions of various orbits. As in typical chaotic systems, one observes that the leading contribution should arise from points in the close vicinity of periodic orbits. From this analysis an analogy with typical chaotic systems can be drawn. As the structure of the formulae in those systems is the same, it can be expected that for some eigenfunctions only a small number of periodic orbits would contribute appreciably. Of-course this cannot be the common case, since usually many periodic orbits contribute, giving rise to elaborate interference patterns. But, for the few eigenfunctions for which this is not the case, scars should be prominent. The question of the existence of scars in the systems studied here is addressed in the next section.

\setcounter{equation}{0}\def\theequation{\arabic{section}.\arabic{equation}}
\section{A search for scars}\label{Sec_PS_Yp}
In this section, the possibility of finding scars, as usually understood by physicists, namely eigenfunctions that are appreciably large on unstable periodic orbits, will be explored The starting point are analytic calculations similar to those of \cite{Agam94a}, for a billiard with periodic boundary conditions. Then, corresponding results are studied for a hard wall billiard. Finally these are tested numerically. 

Scarring of a wavefunction in a quantum system, means that the corresponding probability density is considerably larger than average near some unstable periodic orbit of the corresponding classical chaotic system. Agam and Fishman \cite{Agam94a} suggested to quantify this concept and proposed to test whether an eigenfunction is scarred or not. They defined $Y_\po(E_\alpha)$ as the amount of wavefunction density in a narrow tube in phase space, with a large enough cross section around the orbit $\po$, in excess of the mean value. The term ``large enough'' means that the tube has to be wide enough on the scale of oscillations as the distance from the periodic orbit is increased (governed by $\hbar$). This tube cannot be too wide on the classical length scale if it is to encompass one short periodic orbit only. Using the semiclassical approximation for eigenfunctions (after the integral over the momentum coordinates is performed), the ``scar weight'' is given by (see AF 5.4):
\beq\label{crittube}
Y_\po(E_\alpha)=\Int_{tube_{\po}}d\mu(\zeta)\ABS{ \psi_{\alpha}(\zeta)}^{2}_{po}
~,
\eeq
where $\ABS{ \psi_{\alpha}(\zeta)}^{2}_{po}$ is the contribution of the periodic orbits to the eigenfunction density. Eigenfunctions are normalized, therefore, if the number obtained is large (on a scale of $1$) when compared to integrals over tubes around other periodic orbits, then a considerable fraction of the eigenfunction density lies in this tube and the eigenfunction should be scarred. If, on the other hand, the number is small, the eigenfunction should not be scarred. Agam and Fishman tested this criterion for a truncated hyperbola billiard.

The effect of taking the narrow tube around the periodic orbit $p$ is to remove from \Eq{psiOsc} all elements with large $d_{p.o.}^{\perp}$'s. There are two types of such elements: images of $p$, all other periodic orbits (and their images). This means that the tube removes from \Eq{psiOsc} all of the contributions that do not appear in \Eq{psi-O-bill}. Therefore, the analytic results derived here concur with those of  Agam and Fishman \cite{Agam94a} completely. Namely, if the tube is narrow enough, then the contributions from periodic orbits other than $p$ would generally be negligible. The reason is that, because periodic orbits other than $p$ typically have large $d_{p.o.}^{\perp}$, the phases in \Eq{psiOsc} (as can be seen clearly in \EQ{psi_d_small}) oscillate wildly. On the other hand, there is a narrow region around $p$ in which $d_{p.o.}^{\perp}$ is very small, so that the phase in \Eq{psiOsc} is stationary, and therefore $p$ is dominant.

Instead of integrating over a complicated tube that encompasses many segments of geodesics in the billiard, the symmetry of the eigenfunction may be used to simplify it. Since the eigenfunction is invariant to group isometries it is the same in all the tiles obtained by them. Therefore, a tube of length $L_p$ that extends from the billiard into nearby tiles may be constructed, just like one constructs a periodic orbit, in the following manner:
\begin{enumerate}
	\item Construct a tube around one of the periodic orbit segments in the billiard;
	\item Continue the construction into the next tile the segment continues into;
	\item Repeat this process in all the copies the orbit visits during one period.
\end{enumerate}
Therefore, restricting the integration to a narrow tube surrounding the invariant geodesic of $p$ picks out of \Eq{psiOsc} the contribution of two transformations, $p$ and $p^{-1}$, provided that the tube is narrow enough. The reason for this is the behaviour of the function $\Delta^{(p',h)}_{\mu}(\zeta,k)$ of \EQ{Delta_phmu_approx} (here $p'$ stands for any primitive element from the group) far away from invariant geodesic of $hp'h^{-1}$. This function involves a phase that causes it to oscillate wildly during the integration over the width of the tube. Only for the invariant geodesic of $p$ (namely for the group elements $p$ and $p^{-1}$) can this phase be stationary throughout the length of the tube. This means that the contribution of $p$ and $p^{-1}$ is dominant. This argument is, of-course, the same argument for typical chaotic systems that Agam and Fishman \cite{Agam94a} used.

The resulting scar weight:
\begin{eqnarray}\label{Ypdef}
	Y_p(E_\alpha)&\sim&2\Int_{tube_p}d\mu(\zeta)
\frac{1}{2\pi\Delta'(E_\alpha)}  \Im\Bigg\{ \bar{\Delta}^{(p,I)}(\zeta,k_\alpha) \Bigg\}
~,
\end{eqnarray}
where $p$ is the isometry whose invariant geodesic was chosen as the axis of the tube, and $I$ is the identity. The factor 2 arises because the contribution of $p^{-1}$ is exactly equal to that of $p$.

It is always possible to use isometries to transform to coordinates in which $p$ is diagonal. In such coordinates the invariant geodesic of $p$ is the positive imaginary axis in the Poincar\'e half-plane.  Another isometric coordinate transformation may also be performed, so that the invariant geodesic section under consideration extends from $1$ to $e^{L_p}$.

Now \Eq{Ypdef} reads:
\begin{eqnarray}\label{Yp2}
	Y_p(E_\alpha)&=&\frac{1}{\pi\Delta'(E_\alpha)}  \Im\Bigg\{\sqrt{\frac{\pi}{2ik_\alpha}} \sum_\mu e^{-i\pi\smoothN (E_\alpha)+ik_\alpha L_{\mu,p}}\sum_{j=0}^{j_{p}} \bar{C}_\mu^{(p,j)} {\cal I}_{\mu,j}(k_\alpha)\Bigg\}
~,
\eeqa
where:
\beqa
{\cal I}_{\mu,j} (k)&\equiv& \Int_1^{e^{L_p}} \frac{dy}{y^2}\Int_{-y\sqrt{\cosh{w}-1}}^{y\sqrt{\cosh{w}-1}} dx  \Bigg\{\frac{e^{ik\left( d_{p^{j+1}}\left(x+iy\right)-(j+1)L_p\right)} }{\sqrt{\sinh{d_{p^{j+1}}\left(x+iy\right)}}}
\times\Lf&&\times
\Erfc\left[\sqrt{\frac{k}{2Q^2(K,k)}}\left(L_{\mu,p}-\frac{k{\cal A}}2+\ln\cosh^{2}{d^\perp_p (x+iy)}\right)\right]\Bigg\}
~,
\eeqa
where $w$ is the width of the tube assumed narrow enough to capture only one periodic orbit. Note, that a tube around a geodesic is defined as the geometrical location of all points of constant perpendicular distance ($d^{\perp}$) from it.  In the coordinates in which $p$ is diagonal, $\Cp$ is very simple (see \Fig{fig_perp}):
\beq
\label{perp_dist}
	\Cp=\cosh^2{d_{\zeta,i\ABS{\zeta}}}=1+\left(\frac{x}{y}\right)^2
\eeq
and depends only on the variable:
\beq
X^2=\frac{x^2}{y^2}~.
\eeq
Due to \EQ{beardoneven}, $\cosh{d_{p^{j+1}}\left(\zeta\right)}$ depends only on this variable too. (This result is correct for odd boosts too, due to \EQ{beardonodd}.) 

As the tube is very narrow one may use the approximations \Eq{Qapprox} and \Eq{dperpapprox}.  Changing variables $x\rightarrow X=\frac{x}{y}$ one obtains:
\beqa\label{IntI}
{\cal I}_{\mu,j} (k)&\sim& \frac{L_p}{\sqrt{\sinh{(j+1)L_p}}} \Int_{-\sqrt{\cosh{w}-1}}^{\sqrt{\cosh{w}-1}} dx  \Bigg\{e^{ik\tanh\left(\frac{j+1}{2}L_p\right)X^2}
\times\Lf&&\times
\Erfc\left[\sqrt{\frac{k}{2Q^2(K,k)}}\left(L_{\mu,p}-\frac{k{\cal A}}2+X^2\right)\right]\Bigg\}
\as{k}\infty~.
\eeqa
The integrand of \Eq{IntI} contains an exponent with a rapidly changing phase, so the method of stationary phase may be used. Since most of the contribution to such  integrals comes from an exponentially narrow region around $X=0$, then if the tube was wide enough on the oscillations' scale ($\frac{1}{\sqrt{k}}$), the integration limits may be taken as $\pm\infty$ without introducing a large error. In addition,  the complementary error function hardly changes in the contributing region, so one may approximate it by it's value at $X=0$, \ie on the the periodic orbit.  One then obtains:
\beq\label{Ifinal}
{\cal I}_{\mu,j}(k) \sim \frac{L_p}{\sinh{\frac{(j+1)}{2}L_p}}\sqrt{\frac{i\pi}{2k}} 
\Erfc\left[\sqrt{\frac{k}{2Q^2(K,k)}}\xi_{\mu,p}(k)\right]\Bigg\}
\as{k}\infty~,
\eeq
where $\xi_{\mu,p}(k)=\left(L_{\mu,p}-\frac{k{\cal A}}2\right)$. Inserting \Eq{Ifinal} into \Eq{Yp2}, one obtains:
\begin{eqnarray}\label{Yp}
Y_p(E_\alpha)&\sim&\frac{L_p}{2k_\alpha\Delta'(E_\alpha)}  \Im\Bigg\{ \sum_\mu e^{-i\pi\smoothN (E_\alpha)+ik_{\alpha}L_{\mu,p}}
\times\Lf&&\times
\sum_{j=0}^{j_{p}}\frac{ \bar{C}_\mu^{(p,j)} }{\sinh{\frac{(j+1)}{2}L_p}}
\Erfc\left[\sqrt{\frac{k_\alpha}{2Q^2(K,k_\alpha)}}\xi_{\mu,p}(k_\alpha)\right]\Bigg\}
\as{k_\alpha}\infty~.
\end{eqnarray}

\subsection{The weight of scars for hard wall billiards}
Using the the same definition for the scar weights \Eq{crittube} that was used before, very similar results are obtained:
\begin{eqnarray}\label{Ypbill}
Y_\po(E_\alpha)&\sim&\frac{b_\po L_\po}{2k_\alpha\Delta'(E_\alpha)}  \Im\Bigg\{ \sum_\mu e^{-i\pi\smoothN (E_\alpha)+ik_{\alpha}L_{\mu,\po}}
\times\Lf&&\times
\sum_{j=0}^{j_{\po}}\frac{ \bar{C}_\mu^{(\po,j)} }{\sinh{\frac{(j+1)}{2}L_\po}}
\Erfc\left[\sqrt{\frac{k_\alpha}{2Q^2(K,k_\alpha)}}\xi_{\mu,\po}(k_\alpha)\right]\Bigg\}
\as{k_\alpha}\infty~,
\end{eqnarray}
where 
\beq
b_\po=\left\{\begin{array}{cc} \half& \mbox{$\po$ self-retracing}\\ 1& \mbox{else}\end{array}\right. 
\eeq
The only differences in the derivation and in the final result are numerical prefactors.  For singular periodic orbits an additional factor of half is required, because they run along an edge and therefore only half of the tube around them is really in the billiard. This factor is compensated  by the factor of two associated with the existence of an odd counterpart to every even singular primitive periodic orbit. The factors $b_\po$ result from the fact that if periodic orbit is self-retracing, than $\po^{-1}$ is in the conjugacy class of $\po$ and therefore the prefactor two that appears in  \Eq{Ypdef} is missing.

\subsection{A numerical search for scars}
In order to check for scars on some periodic orbits of the billiard defined in \Fig{tile}, the scar weight of \EQ{Ypbill} was calculated on the roots of the semiclassical approximation of the spectral determinant (see \Sec{sec_spec_det_sel}), that are the semiclassical approximation to the eigenenergies. The results for the four shortest primitive periodic orbits are presented in \Fig{figYp1234}. Also presented in the figure are the locations of the energies that admit standing waves on the respective periodic orbits. These are the energies satisfying the Bohr-Sommerfeld quantization rule. For billiards with Dirichlet boundary conditions it reads:
\beq\label{mybohr}
k_n=\frac{2\pi}{L_\po}\left(n+\half n_{b} \right)~~\left(=\sqrt{E_n-\qrt}\right)
~,
\eeq
where $L_\po$ is the length of a periodic orbit and $n_{b}$ is the number of times it bounces off the walls during one period.

The sum rule \cite{Agam94a}:
\beq\label{sumrule}
\sum_{\po}Y_\po(E_\alpha) \approx 1-\frac{\bar{\Delta_i}(E_\alpha)}{\Delta'(E_\alpha)}\pi\bar\rho(E_\alpha)
\eeq
was found to be satisfied to a very good approximation in every case studied. In \EQ{sumrule}, $\bar\rho(E_\alpha)$ is given by the derivative of $\smoothN (E_\alpha)$ \Eq{weyl_b} with respect to $E$. \EQ{sumrule} is the expected result for $k_\alpha\rightarrow\infty$, because the stationary phase limit was used to calculate the scar weight from it's definition, \Eq{crittube}, \ie integrating over a narrow tube and integrating over the whole billiard is the same as $k_\alpha\rightarrow\infty$. Using the fact that the eigenfunction is normalized, obtaining \Eq{sumrule} is straight forward.

The general features of the scar weight, $Y_\po(E_\alpha)$, on various periodic orbits for the eigenenergies as presented in \Fig{figYp1234} agree with the general features found in \cite{Agam94a}. Prominent among these is that the local maxima tend to be located at eigenenergies close to energies that admit standing waves, given by \Eq{mybohr}. 

The main difference between \Fig{figYp1234} and the figures presented in \cite{Agam94a}, is that it seems that the functions found in the present work are not as smooth as in the previous work. As a result of this, it is difficult to find an eigenenergy for which $Y_\po(E_\alpha)$ is negligible for all $\po$'s but for a few, because the scar weights of various periodic orbit fluctuate strongly. The reason for the erraticness is not clear at this stage. As a result, it is difficult to implement Agam and Fishman's criterion for scars here.

In order for an eigenfunction to be scarred it must have a pronounced contribution from only a few periodic orbits. Therefore the scarred eigenfunctions should be the ones for which the scar criterion is appreciable on one or two periodic orbits only. Agam and Fishman \cite{Agam94a} found this to be in very good agreement with the features observed in the eigenfunction densities they calculated numerically, for a truncated hyperbola billiard.

In addition to this, one must bear in mind that in order for a scar to be prominent it must be concentrated in a narrow tube around a periodic orbit, where ``narrow'' means narrow with respect to relevant classical linear dimensions in the billiard (for example, the distance between short periodic orbits). The width of the tube in which one may expect to see a scar is given by the scale on which the oscillations, as a function of the distance from the periodic orbit, occur. According to \Eq{psi_d_small}, this scale is approximately given by:
\beq
l_{osc}\approx\frac{1}{\sqrt{k_\alpha}}
~.
\eeq
The highest energy for which the results here were trustworthy (from the numerical calculation of $\Delta(E)$) was approximately
\beq
E_\alpha\approx5000~,
\eeq
that corresponds to  
\beq
l_{osc}\approx0.12
~.
\eeq
For comparison, the distance between the point the shortest periodic orbit hits the side $b$ and the point on that side where the second shortest periodic orbit hits, is $0.16$ (see \Fig{figShorts}). This means that even at the highest eigenenergy for which the scar criterion could be applied with any reliability, within our numerical limitations, the tube where one expects to see scars is too wide for resolving individual periodic orbits. Nevertheless, a numerical search for scars was initiated.

There are two ways to go about such a search. The first, which yielded good results in the past \cite{Agam94a}, is to try to use the scar weight graphs (see \Fig{figYp1234}) to predict whether or not a prominent scar exists in some eigenfunction. The second method will be described below. The first was used here, without decisive results. When comparing the predictions with the figures of the numerically found eigenfunctions, sometimes scar-like features were observed where they were expected and sometimes they were not. Some figures of eigenfunction densities are presented in \Fig{waves84689091} and \Fig{waves9293}. In principle these figures are slightly misleading, because the metric \Eq{Disk_dmu} is not constant. But, as the billiard is not very extended, as can be seen in \Fig{tile} and \ref{figShorts}, the aberration is not large, because:
\beqar
d\mu&=&\frac{4r~drd\phi}{\left(1-r^2\right)^2}= 4\left(1+O[r^2]\right)rdrd\phi 
~.
\eeqar
As the shortest periodic orbits lie in a region with $r\le\qrt$ approximately, the deviation of the invariant measure from the Euclidean invariant measure is negligible.
  
For example, according to \Fig{figYp1234} eigenfunction \#$84$ should be scarred along the second shortest periodic orbit, while eigenfunction \#$68$  and eigenfunction \#$90$ should be scarred along the fourth shortest periodic orbit. In \Fig{waves84689091} one can see that indeed there are some high peaks of these eigenfunctions along the said periodic orbits. On the other hand, according to \Fig{figYp1234}, eigenfunction \#$91$ should not be very scarred along the fourth shortest periodic orbit., while in \Fig{waves84689091} one can see that it does have some high peaks along that periodic orbit. Another example is eigenfunction \#$92$ which, according to \Fig{figYp1234}, should have exhibited a scar along the shortest periodic orbit and does not, as can be seen in \Fig{waves9293}. In the same figure one can see that the next level, \#$93$, does show a scar on that periodic orbit, as the scar criterion predicts.

At this point it is worth mentioning that the scars observed in \cite{Agam94a,Heller84} and more recently in \cite{Klakow96} tended to be more pronounced than the scar-like features observed in the present work.

One may also choose an opposite strategy in order to check the viability of the scar criterion. This strategy is faulty, as will be explained shortly. In this method one looks for scar-like features in eigenfunctions, and then checks the scar criterion graphs for a correspondence. There were several discrepancies, one of which was mentioned above, namely eigenfunction \#$91$ (see \Fig{waves84689091}) which has a prominent scar-like feature on part of the fourth shortest periodic orbit. This strategy is misleading because there are many complicated structures in high energy eigenfunctions in a billiard, as they are the result of an elaborate diffraction pattern. This means that a scar-like structure may have nothing to do with a periodic orbit, as it may be just an accidental pattern. The first strategy, described above, filtered out some of these scar look-alikes automatically, because a large scar weight indicates that the structure observed may be related to a periodic orbit. 

The reason why one expects scars in the first place, despite claims to the contrary \cite{Aurich91,Hejhal92,Aurich93}, is based on the observation that many features of the special class of systems under study here are similar to those of typical chaotic systems (for example \EQ{psi_d_small}). Since Agam and Fishman \cite{Agam94a} found scars in eigenfunctions whose energy admitted standing waves, one is led to expect the same here. As many prominent scars were not found in the system under numerical study, although the scar weight graphs suggest that scars should be present where standing waves occur, there may be important differences between the systems studied here and typical ones.

An important difference that may be the cause for the scarcity of scars in the eigenfunctions examined is the very high topological entropy of billiards on the pseudo-sphere. For example, in the hyperbola billiard, a derivative of which Agam and Fishman \cite{Agam94a} used, the number of periodic orbits up to length $d$ is \cite{Sieber90}:
\beq
N(d)\sim \frac{e^{0.6d}}{0.6d} \as{d}{\infty}
~.
\eeq
This is much less than in tiling billiards on the pseudo-sphere, as can be seen in \Eq{exp_prolif}. This leads to the following consequences:
\begin{enumerate}
	\item It is difficult to find all the primitive periodic orbits up to some length;\nopagebreak
	\item The periodic orbits are denser than in the typical case, so shorter wavelengths may be required to observe scars clearly.
\end{enumerate}

Evidently, it would have been advantageous to study states with higher energy, that corresponds to shorter wavelength, than was done in the present work. Unfortunately computational limitations prevent this, leaving the numerical results inconclusive, as to the existence of scars.

\setcounter{equation}{0}\def\theequation{\arabic{section}.\arabic{equation}}
\section{Summary and conclusions}\label{summary}
In this work an exact expression for the eigenfunctions of special systems, whose classical dynamics are chaotic, was obtained. These systems, tiling billiards on the pseudo-sphere, are special in that many formulae, that  for typical chaotic systems are correct only in the leading semiclassical approximation, are exact for them. Specific differences between the exact expressions, as well as the approximate ones, and those used for general chaotic systems were identified explicitly. 

The main result of this work is \EQ{green_ratio}, an exact expression for (the diagonal part of) Green's function, in the form:
\beq\label{ratioLast}
G_E(\zeta,\zeta)=\frac{{\cal N}(\zeta,E)}{\Delta(E)}
\eeq
whose residues are the squares of the exact eigenfunctions of the Hamiltonian \Eq{exact_eig}. This result is a manifestation of Fredholm theory \cite{Fishman96} for the systems studied here, because that theory predicts that Green's function in general can be expressed as such a ratio. The special feature of tiling billiards on the pseudo-sphere is that both numerator and denominator can be expressed in terms  periodic orbits, {\em without the introduction of any approximation}. It would be interesting to study the exact connection between the results presented here and those of Fredholm theory. Such a connection would supply a recipe for reordering the problematic sums in the numerator and the denominator to ensure rapid convergence to the {\em exact} result. 

Summation over repetitions, leads to the form \Eq{ratioLast} of the Green function and to the appearance of the spectral determinant in the denominator. Therefore repetitions directly determine the spectrum. Any specific zero is not related to a specific factor in the product of terms of the form of \Eq{Delta_prod}, but rather to a zero of the the product that does not converge absolutely. This mechanism has been found in earlier work in the framework of the semiclassical approximation \cite{Agam93} and of Fredholm theory \cite{Fishman96}, but here it is established explicitly in the framework of the exact theory.

It was shown that there are three different kinds of contributions to the exact ${{\cal N}(\zeta,E)}$. The first of these is associated with the non-periodic orbits of the system, the second with the periodic orbits, and the third with their images. The first two contributions appear in the leading semiclassical approximation for Green's function (and for eigenfunctions) of typical chaotic systems.

The third contribution, that of images of periodic orbits, can be given the following interpretation. Such a contribution can be associated with a non-periodic closed orbit, whose length is $d_{h\po h^{-1}}(\zeta)$. In typical chaotic systems this contribution is approximated by nearby periodic orbits, but in a tiling billiard on the pseudo-sphere, the fact that ${h\po h^{-1}}$ belongs to a conjugacy class that corresponds to a specific periodic orbit, provides an exact recipe for accounting for it's contribution. In general, such a correspondence cannot be found with the standard semiclassical approximation methods, because the closed orbit ${h\po h^{-1}}$ is associated, with need not lie close to the periodic orbit to which $\po$ is related.

In the general result obtained for chaotic systems, contributions are taken only from the close vicinity of periodic orbits, assuming that contributions from distant ones cancel due to destructive interference \cite{Bogomolny88,Berry89a}. Here, such an assumption was not required, since the contributions from all periodic orbits can be calculated exactly. If further approximations, namely an expansion around periodic orbits, are made, the expressions reduce to those found for general chaotic systems. For general chaotic systems, the results can be obtained only in the framework of such an approximation. For the systems that are studied here, the only correction to the semiclassical results are the contributions from images of periodic orbits, namely orbits outside the billiard.

The second result obtained is a semiclassically approximate expression, resulting from an effective truncation of the infinite sums over periodic orbits of the exact ${{\cal N}(\zeta,E)}$ and ${\Delta(E)}$ by the Berry-Keating resummation \cite{Berry92}. The truncated expressions were compared with previous results for typical chaotic systems \cite{Agam93}. As the previous results were derived assuming that the dominant contributions arise from the close vicinity of periodic orbits, the same limit was checked for the results obtained here. The two expressions turned out to be identical.

The resummed expression for eigenfunctions of hard wall billiards was checked numerically for the two lowest states, and reasonable, although not perfect, results were obtained.  In principle, higher orders of the semiclassical approximation can be derived for the systems studied. Inclusion of these in the numerics should yield better results for the low eigenstates checked.

In addition, a criterion for scars \cite{Agam94a} was derived for the chaotic systems at hand. It was tested numerically to check whether scars exist in the systems studied. Inconclusive results were obtained, as the energy attainable in this work was not high enough, due to computational limitations. In order to resolve the question of scars, better numerics are required. One way to go about this would be to work with arithmetic billiards, although they are not generic, as for them there exist systematic methods for obtaining long lists of periodic orbit lengths, that are complete. This should enable the study of states with high energies.

Throughout this work the nature of the exact results obtained for tiling billiards on the pseudo-sphere was found to be very similar to results that are only semiclassical approximations for typical chaotic systems. Therefore, tiling billiards on the pseudo-sphere may be useful as a testing ground for approximate results concerning typical chaotic systems.

{\Large \bf Acknowledgments}

We are indebted to Oded Agam for very useful discussions and suggestions. We have also benefited from discussions with Bertrand Georgeot, Eugene Gutkin, Romanas Narevich, Ze\'ev Rudnick and Denis Ullmo. This work was supported in part by the U.S.-Israel Binational Science Foundation (B.S.F.), the fund for the Promotion of Research at the Technion and the Minerva Center for Nonlinear Physics of Complex Systems.

\setcounter{section}{0}\def\thesection{Appendix \Alph{section}}
	\setcounter{equation}{0}\def\theequation{\Alph{section}.\arabic{equation}}
\section{The derivation of Eq.~\protect\ref{delta_i_ophir}}\label{Msum}
In this appendix the sum over $m$ in $e^{\frac i\hbar{\bf{\widetilde X} R_{\po} X}}\sum\limits_{m,\mu}C_{\mu}^{({\po},m)} g_m\left[\frac i\hbar{\bf{\widetilde X} R_{\po} X}\right]$ of \Eq{delta_i_oded} will be performed, leading to \Eq{delta_i_ophir}. Since the sum over $\mu$ and $m$ converges absolutely, the order of terms can be changed. In particular, exchanging the summation over composite orbits with the summation over $m$ and using \EQ{Cmupm}, one finds:
\beqar
e^{\frac i\hbar{\bf{\widetilde X} R_{\po} X}}\sum_{m=0}^{\infty} C_{\mu}^{({\po},m)} g_m\left[\frac i\hbar{\bf{\widetilde X} R_{\po} X}\right] &=& C_{\mu-{\po}}(-1)^{j_{\po}}e^{-\half(j_{\po}+1)u_{\po}-i(j_{\po}+1)\gamma_{\po}}\times\\
&&\!\!\!\!\!\!\!\!\!\!\!\!\!\!\!\!\!\!\!\!\!\!\!\!\!\!\!\!\!\!\!\!\!\!\!\!\!\!\!\!\!\!\!\!\!\!\!\!
\times
\sum_{m=0}^{\infty} g_m\left[\frac i\hbar{\bf{\widetilde X} R_{\po} X}\right]e^{\frac i\hbar{\bf{\widetilde X} R_{\po} X}}\sum_{j=0}^{j_{\po}}(-1)^je^{-m(j+1)u_{\po}}d_{j_{\po}-j}\left(e^{-u_{\po}}\right)\nonumber
~.
\eeqar
Exchanging summation order again, one obtains:
\beqar
e^{\frac i\hbar{\bf{\widetilde X} R_{\po} X}}\sum_{m=0}^{\infty} C_{\mu}^{({\po},m)} g_m\left[\frac i\hbar{\bf{\widetilde X} R_{\po} X}\right] &=& C_{\mu-{\po}}(-1)^{j_{\po}}e^{-\half(j_{\po}+1)u_{\po}-i(j_{\po}+1)\gamma_{\po}}\times\\
&&\!\!\!\!\!\!\!\!\!\!\!\!\!\!\!\!\!\!\!\!\!\!\!\!\!\!\!\!\!\!\!\!\!\!\!\!\!\!\!\!\!\!\!\!\!\!\!\!
\times
\sum_{j=0}^{j_{\po}}(-1)^j d_{j_{\po}-j}\left(e^{-u_{\po}}\right)\Bigg\{\sum_{m=0}^{\infty} g_m\left[\frac i\hbar{\bf{\widetilde X} R_{\po} X}\right]e^{\frac i\hbar{\bf{\widetilde X} R_{\po} X}}e^{-m(j+1)u_{\po}}\Bigg\}\nonumber
~.
\eeqar
Comparing the curly bracketed term with \Eq{taylor} it is easy to see that:
\beqar
\Bigg\{\sum_{m=0}^{\infty} g_m\left[\frac i\hbar{\bf{\widetilde X} R_{\po} X}\right]e^{\frac i\hbar{\bf{\widetilde X} R_{\po} X}}e^{-m(j+1)u_{\po}}\Bigg\}\equiv\frac{\exp\left[  
\frac{i}{\hbar} \widetilde{{\bf X}} J \frac
{M_{{\po}}^{j+1}-I}{M_{{\po}}^{j+1}+I} {\bf X}\right]}{\sqrt{\det (M_{{\po}}^{j+1}+I)}} e^{\half(j+1)u_{\po}}
~,
\eeqar
so that in conclusion, since in two dimensions (see AF 3.11 \& 3.13):
\beqar
\frac{\exp\left[  
\frac{i}{\hbar} \widetilde{{\bf X}} J \frac
{M_{{\po}}^{j+1}-I}{M_{{\po}}^{j+1}+I} {\bf X}\right]}{\sqrt{\det (M_{{\po}}^{j+1}+I)}}=\frac{\exp\left[  
\frac{i}{\hbar} {\bf{\widetilde X} R_{\po} X}~\tanh{\frac{j+1}2u_{{\po}}}\right]}{2\cosh{\frac{j+1}2u_{\po}}}
~,
\eeqar
one obtains\newpage
\beqar
\sum_{m=0}^{\infty}  C_{\mu}^{({\po},m)}g_m&&\left[\frac i\hbar{\bf{\widetilde X} R_{\po} X}\right]e^{\frac i\hbar{\bf{\widetilde X} R_{\po} X}} = C_{\mu-{\po}}(-1)^{j_{\po}}e^{-\half(j_{\po}+1)u_{\po}-i(j_{\po}+1)\gamma_{\po}}
\times\Lf&&\times
\sum_{j=0}^{j_{\po}}(-1)^j e^{\half(j+1)u_{\po}} d_{j_{\po}-j}\left(e^{-u_{\po}}\right)\frac{\exp\left[  
\frac{i}{\hbar} {\bf{\widetilde X} R_{\po} X}~\tanh{\frac{j+1}2u_{{\po}}}\right]}{2\cosh{\frac{j+1}2u_{\po}}}
=\Lf&&=
\sum_{j=0}^{j_{\po}}\frac{\bar{C}_\mu^{({\po},j)}}{2\cosh{\frac{j+1}2u_{\po}}}
\exp\left[\frac{i}{\hbar} {\bf{\widetilde X} R_{\po} X}~\tanh{\frac{j+1}2u_{{\po}}}\right]
~.
\eeqar
$\bar{C}_\mu^{({\po},j)}$ was defined in \Eq{def_Cmupj}. Substitution of the last equation in \Eq{delta_i_oded} leads to \Eq{delta_i_ophir}.

	\setcounter{equation}{0}\def\theequation{\Alph{section}.\arabic{equation}}
\section{Writing Legendre functions as sums over $e^{-u_\po}$}\label{Q}
To prove \Eq{legendre-g2} one starts with the integral representation of the relevant Legendre function \cite{GR80}:
\begin{eqnarray*}
\label{intQrep}
Q_{-\half-ik}\left[\cosh{d}\right]=\frac{e^{ikd}}{\sqrt{2}}\Int_{0}^{\infty}\frac{e^{iky}dy}{\sqrt{\cosh(d+y)-\cosh(d)}}~~\mbox{for}~d>0
~,
\end{eqnarray*}
which can be rewritten:
\begin{eqnarray*}
Q_{-\half-ik}\left[\cosh{d}\right]=e^{(ik-\half)d}\Int_{0}^{\infty} dy\frac{e^{iky}}{\sqrt{e^{y}-1}}\sum_{n=0}^{\infty}\frac{(2n-1)!!}{(2n)!!}\left(e^{-y-2d}\right)^{n}
~.
\end{eqnarray*}
The summation and integration order can be reversed, because the integral and the sum are absolutely convergent:
\begin{eqnarray*}
Q_{-\half-ik}\left[\cosh{d}\right]=\sum_{n=0}^{\infty}\frac{(2n-1)!!}{(2n)!!}e^{(ik-\half-2n)d}\Int_{0}^{\infty} dy\frac{e^{-(n-ik)y}}{\sqrt{e^{y}-1}}
~.
\end{eqnarray*}
The last integral is known \cite{GR80}, and one obtains:
\begin{eqnarray}\label{Qsum_e^d}
Q_{-\half-ik}\left[\cosh{d}\right]=\sum_{n=0}^{\infty}\frac{(2n-1)!!}{(2n)!!}B\left[n+\half-ik,\half\right]e^{(ik-\half-2n)d}
~.
\end{eqnarray}
Since the $d$'s to be used in the above expression are the $\dg$'s of \EQ{beardoneven}, one can write, after some straight-forward algebra:
\begin{eqnarray*}\label{e^d}
e&&^{d_{g}(\zeta)}=\Cg e^{L_{g}} \cdot f\left[x,d^{\perp}_{g}(\zeta)\right]_{\big|_{x=e^{-L_{g}}}}
~,
\end{eqnarray*}
where
\beqar
f\left[x,d^{\perp}_{g}(\zeta)\right]\equiv(1-x)^{2}\half\left[1+\sqrt{1+\frac{4}{\Cg} \frac{x}{(1-x)^{2}}} + \frac{2}{\Cg} \frac{x}{(1-x)^{2}} \right]
~.
\eeqar
Now $\left[e^{d_{g}(\zeta)}\right]^{ik-\half-2n}$ can be expanded in powers of $x=e^{-L_{g}}$:
\beqa\label{e^d_expansion}
\left[e^{d_{g}(\zeta)}\right]^{ik-\half-2n}&=&\left[\frac{e^{-L_g}}{\Cg}\right]^{\half+2n-ik}\cdot \Big\{f\left[x,d^{\perp}_{g}(\zeta)\right]\Big\}^{ik-\half-2n}=
\Lf&&\hspace{2cm}=
\sum_{m=0}^{\infty}a_m\left[\frac{1}{\Cg},ik-\half-2n\right] e^{-mL_{g}}
~,
\eeqa
where the first few $a_m[x,y]$'s are:
\begin{eqnarray*}\label{a_m}
a_0\left[x,y\right]&=&1 ~,\\
a_1\left[x,y\right]&=&-2y+2yx
~.
\end{eqnarray*}

Since the $a_m[x,y]$'s are polynomials of $y$, one may write:
\beqar
a_m[x,y]=a_m[x,\partial_\eta]e^{\eta y}~_{\big|_{\eta=0}}
~,
\eeqar
to bring \Eq{e^d_expansion} to a more convenient form:
\begin{eqnarray}\label{e^d_expansion2}
\left[e^{d_{g}(\zeta)}\right]^{ik-\half-2n}&=&
\sum_{m=0}^{\infty}a_m\!\left[\frac{1}{\Cg},\partial \eta\right]\left(\frac{e^{-\eta}}{\Cg}\right)^{\half+2n-ik}_{ \big|_{\eta=0}}\hspace{-1cm}\times e^{ikL_{g}}e^{-(\half+m+2n)L_{g}}
~.
\end{eqnarray}
Inserting \Eq{e^d_expansion2} into \Eq{Qsum_e^d} one obtains:
\begin{eqnarray}\label{legendre-g}
Q_{-\half-ik}\left[\cosh{d_{g}(\zeta)}\right]=
\sum_{n=0}^{\infty}\sum_{m=0}^{\infty}\frac{(2n-1)!!}{(2n)!!}B\left[n+\half-ik,\half\right]  &&\times \nonumber \\
\times a_m\left[\frac{1}{\Cg},\partial \eta\right]\left(\frac{e^{-\eta}}{\Cg}\right)^{\half+2n-ik}_{  \big|_{\eta=0}}\hspace{-1cm}\times &&e^{ikL_{g}}e^{-(\half+m+2n)L_{g}}
~.
\end{eqnarray}
Changing summation variables in \Eq{legendre-g}:
\beqa\label{sumExchange}
\sum_{n=0}^{\infty}\sum_{m=0}^{\infty} A_n B_m \stackrel{m'=m+2n}{\longrightarrow} \sum_{n=0}^{\infty}\sum_{m'=2n}^{\infty} A_n B_{m'-2n} \longrightarrow \sum_{m'=0}^{\infty}\sum_{n=o}^{\left[\frac{m'}{2}\right]} A_n B_{m'-2n} 
~,
\eeqa
leads to \EQ{legendre-g2}.

{\large \bf The alterations for odd boosts}
\nopagebreak

In deriving \Eq{Nosc}, \Eq{beardoneven} was relied upon to obtain an expansion of $e^{\dg}$ in powers of $e^{-L_p}$, when the group included only even boosts. For odd boosts \Eq{beardonodd} will be used, leading to:
\begin{eqnarray}
\label{odde^d}
e^{d_{g}}{}_{\big|_{g\in odd}}&=&\Cg e^{L_{g}} \times \Lf
&&\times(1+x)^{2}
\half\left[1+\sqrt{1-\frac{4}{\Cg} \frac{x}{(1+x)^{2}}}  - \frac{2}{\Cg} \frac{x}{(1+x)^{2}} \right] {}_{\big|_{x=e^{-L_{g}}}}
\end{eqnarray}
One can see that the only difference between \Eq{odde^d} and \Eq{e^d}  is the change $x\rightarrow -x$. This means that when \Eq{odde^d} is expanded in powers of $x$, an expansion similar to \Eq{e^d_expansion} will be obtained, but for a factor $(-1)^m$. Therefore:
\beqar
a_m{}_{|_{g{~odd}}}=(-1)^m~a_m{}_{|_{g{~even}}} ~.
\eeqar
The last equation means that the only alteration that \Eq{legendre-g2} requires for odd boosts is the multiplication of each term in the series by $(-1)^m$. \EQ{legendre-q} and \ref{legendre-s} follow from this immediately.
	\setcounter{equation}{0}\def\theequation{\Alph{section}.\arabic{equation}}
\section{The coefficients $C_{\mu}$ and $C_{\mu}^{(\po,m)}$}\label{C}
The coefficients $C_\mu$ are defined in \Eq{Cmu_berry}. For tiling billiards on the pseudo-sphere they are given by:
\beqa\label{Cmu_Sel}
\hspace{-3cm}C_\mu&=&
\prod_{\{p\}_\mu}(-1)^{j_p}e^{-\half j_p L_p}d_{j_p}\left(e^{-L_p}\right)
\times\Lf&&\times
\prod_{\{q\}_\mu}e^{-\half j_q L_q}d_{j_q}\left(-e^{-L_q}\right)
\times
\prod_{\{s\}_\mu}(-1)^{j_s}e^{-\frac32 j_s L_s}d_{j_s}\left(e^{-2L_s}\right)
~,
\eeqa
\eseq
where the definitions in \Eq{neg_u_p_maslov} were used, and the function $d_{j_p}\left(x\right)$ was defined in \Eq{dj}. Notice that the $C_\mu$'s for the billiards under consideration are real numbers.

The coefficients $C_{\mu}^{({\po},m)}$ ($\po$ stands for $p$, $q$ or $s$) are also all real. Using their definition of \Eq{Cmupm}, and the definitions of \EQ{neg_u_p_maslov}, they are given by:
\bseq
\beqa
C_{\mu}^{({p},m)}&=&C_{\mu-p}(-1)^{j_{p}}e^{-\frac12(j_{p}+1)L_{p}}\sum_{j=0}^{j_{p}}(-1)^je^{-m(j+1)L_{p}}d_{j_{p}-j}\left(e^{-L_{p}}\right)~,\\
C_{\mu}^{({q},m)}&=&C_{\mu-q}e^{-\half(j_{q}+1)L_{q}}\sum_{j=0}^{j_{q}}(-1)^{j(m+1)+m}e^{-m(j+1)L_{q}}d_{j_{q}-j}\left(-e^{-L_{p}}\right)~,\\
C_{\mu}^{({s},m)}&=&C_{\mu-s}(-1)^{j_{s}}e^{-\frac32(j_{s}+1)L_{s}}\sum_{j=0}^{j_{s}}(-1)^je^{-2m(j+1)L_{s}}d_{j_{s}-j}\left(e^{-2L_{s}}\right)
~.
\eeqa
\eseq

	\setcounter{equation}{0}\def\theequation{\Alph{section}.\arabic{equation}}
\section{Resummation of the numerator of Eq.~\protect\ref{green_ratio}}\label{resum}
It is useful to write  ${\cal N}_{po}(\zeta,E)$ of \Eq{Nosc} in a form where the dependence on $k$ is shown explicitly:
\begin{eqnarray}
\label{Nosc2}
{\cal N}_{po}\big(\zeta,E{(k)}\big)&=&
\frac{-1}{2\pi}\sum_{\{p\}}\sum_{m=0}^{\infty}\sum_{h\in \Gamma/\Gamma_{p}}\sum_{\mu}\sum_{n=0}^{\left[\frac{m}{2}\right]}C_{\mu}^{(p,m)}A_{h,n}^{(p,m)}\left[\frac{1}{\Chp},\partial{\eta},\eta\right] \times\nonumber\\
&&\times B\left[n+\half-ik,\half\right]e^{-i\pi\smoothN(E)+ik\bar{L}_{\mu,p}^h(\zeta,\eta)}_{\big|_{\eta=0}}
~,
\end{eqnarray}
where
\beqar
\bar{L}_{\mu,p}^h(\zeta,\eta)\equiv L_{\mu,p}+\ln\left(\Chp\right)+\eta~,
\eeqar
and the definition of ${\cal F}_m\left[k,d^\perp_{h^{-1}ph}(\zeta)\right]$ of \Eq{F} was used to define:
\begin{eqnarray}
\label{Ap}
A_{h,n}^{(p,m)}\left[x,y,z\right]
\equiv
\sum_{n=0}^{\left[\frac{m}{2}\right]}\frac{(2n-1)!!}{(2n)!!}
a_{m-2n}\left[x,y\right]\left(\frac{e^{-z}}{x}\right)^{\half+2n}
~.
\end{eqnarray}
Using \Eq{Nsmooth} and \Eq{Nosc2} in \Eq{NandU} one obtains:
\bseq
\begin{eqnarray}
\label{NumSmoothDef} 
 \bar{\cal N}\big(E{(k)}\big)&\equiv&-\frac{1}{2\pi}\sum_{\mu}\left[U_\mu(k)+U_\mu(-k)\right]~,\\
\label{NumOscDef} 
{\cal N}_{po}\big(\zeta,E{(k)}\big)&\equiv&-\frac{1}{2\pi}\sum_{\{p\}}\sum_{h\in \Gamma/\Gamma_{p}}\sum_{m=0}^{\infty}\sum_{\mu}\left[U_{p,h,m,\mu}(\zeta,k)+U_{p,h,m,\mu}(\zeta,-k)\right]
~,
\end{eqnarray}
\eseq
where:
\bseq
\begin{eqnarray}
\label{Umu}
U_\mu(k)&\equiv&C_{\mu}e^{-i\pi\smoothN (E(k))+ikL_\mu}\frac{1}{2\pi i}
\Int_{\infty+i(1/2+\varepsilon)}^{-\infty+i(1/2+\varepsilon)}
\bigg\{  \frac{dz}{z} \gamma (z,k)    
 \times\nonumber\\&&\hspace{2cm}\times 
Q_{-\half-i(z+k)}\left(1\right)
e^{-i\pi\left(\smoothN (E{(z+k)})-\smoothN (E{(k)})\right)+izL_\mu} \bigg\}
~,
\\
\label{Qphmmu_int}
U_{p,h,m,\mu}(\zeta,k)&\equiv&
\sum_{n=0}^{\left[\frac{m}{2}\right]}C_{\mu}^{(p,m)}A_{h,n}^{(p,m)}\left[\frac{1}{\Chp},\partial{\eta},\eta\right]
e^{-i\pi\smoothN (E(k))+ik\bar{L}_{\mu,p}^h(\zeta,\eta)} 
\times\Lf&&\hspace{0cm}\times 
\frac{1}{2\pi i}\Int_{\infty+i(1/2+\varepsilon)}^{-\infty+i(1/2+\varepsilon)}
\bigg\{\frac{dz}{z} \gamma (z,k)B\left[n+\half-i(z+k),\half\right]
\times\nonumber \\&&\hspace{4cm}\times 
e^{-i\pi\left(\smoothN(E{(z+k)})-\smoothN (E(k))\right)+iz\bar{L}_{\mu,p}^h(\zeta,\eta)}\bigg\}
_{\big|_{\eta=0}}
.
\end{eqnarray}
\eseq
It turns out that: 
\bseq
\begin{eqnarray}
\label{symink1}
U_\mu(-k)&=&{U^{\ast}_\mu(k)}~, \\
\label{symink2}
U_{p,h,m,\mu}(\zeta,-k)&=&{U^{\ast}_{p,h,m,\mu}(\zeta,k)}~.
\end{eqnarray}
\eseq
This is because the $C_{\mu}$'s and $C_{\mu}^{(p,m)}$'s are all real (see \App{C}). In addition \cite{Berry92}:
\beq
\smoothN (E{(-k)})=-\smoothN (E{(k)})
~,
\eeq
and \cite{GR80}:
\bseq
\begin{eqnarray}
\label{integralrepQ}
Q_{-\half-ik}\left[1\right]&=&\frac{1}{\sqrt{2}}\Int_{0}^{\infty}\frac{dt~e^{ikt}}{\sqrt{\cosh{t}-1}}=Q_{-\half+ik}^\ast\left[1\right]~,\\
\label{integralrepB}
B\left[n+1/2-ik,1/2\right]&=&\Int_0^1 dt~\frac{t^{n-1/2-ik}}{\sqrt{1-t}}=B^\ast\left[n+1/2+ik,1/2\right]
~.
\end{eqnarray}
\eseq

Inserting  \Eq{symink1} and \Eq{symink2} into \Eq{NumSmoothDef} and \Eq{NumOscDef} yields the manifestly real expressions:
\bseq
\begin{eqnarray}
\label{NumSmooth1} 
 \bar{\cal N}\big(E{(k)}\big)&=&-\frac{1}{\pi}\sum_{\mu}\Re\Big\{U_\mu(k)\Big\}~,\\
\label{NumOsc1} 
{\cal N}_{po}\big(\zeta,E{(k)}\big)&=&-\frac{1}{\pi}\sum_{\{p\}}\sum_{h\in \Gamma/\Gamma_{p}}\sum_{m=0}^{\infty}\sum_{\mu}\Re\Big\{U_{p,h,m,\mu}(\zeta,k)\Big\}
~.
\end{eqnarray}
\eseq
The next aim is to obtain an approximation for the integrals in \Eq{Umu} and \Eq{Qphmmu_int}. From Weyl's asymptotic series \Eq{weyl_a} one obtains:
\begin{equation}
\label{weyl2}
\pi\left(\smoothN (E{(z+k)})-\smoothN (E(k))\right) \sim \frac{{\cal A}k}{2}z + \frac{{\cal A}}{4}z^2
\as{k}\infty~.
\end{equation}
Notice that the series truncates after $z^2$ because the next terms are all {\em exponentially} small in $k$ \cite{Balazs86}. This fact is specific to billiards with periodic boundary conditions, and will not hold for billiards with reflecting walls. 

Using the integral representations of the special functions in the integrands, \Eq{integralrepQ} and \Eq{integralrepB}, double integrals are encountered. Since the integrals are all absolutely convergent, the order of integration can changed in each case. Then one encounters, in the leading order, an integral that can be evaluated analytically:
\begin{equation}
\label{Erfc_int}
{\cal I}\left[Q^2,\xi\right]\equiv \lim_{\varepsilon\rightarrow0^+}{
\frac{1}{2\pi i}\Int_{\infty+i(1/2+\varepsilon)}^{-\infty+i(1/2+\varepsilon)}
\frac{dz}{z}e^{-\frac{Q^2}{2k}z^2+i\xi z}}=\half\Erfc\left[\sqrt{\frac{k}{2Q^2}}\xi\right]
~.
\end{equation}
Using \Eq{Erfc_int} in \EQ{Umu} and \ref{Qphmmu_int} one finds:
\bseq
\begin{eqnarray}
\label{Umu2}
U_\mu(k)&\sim&C_{\mu}e^{-i\pi\smoothN (E(k))+ikL_\mu}\frac{1}{\sqrt{2}}
\Int_{0}^{\infty}
\frac{dy~e^{iky}}{\sqrt{\cosh{y}-1}}\half\Erfc\left[\sqrt{\frac{k}{2Q^2(K,k)}}\left(\xi_\mu(k)+y\right)\right],
\end{eqnarray}
\begin{eqnarray}
\label{Uphmmu2}
U_{p,h,m,\mu}(\zeta,k)&\sim&
\sum_{n=0}^{\left[\frac{m}{2}\right]}C_{\mu}^{(p,m)}A_{h,n}^{(p,m)}\left[\frac{1}{\Chp},\partial{\eta},\eta\right]
e^{-i\pi\smoothN (E(k))+ik\bar{L}_{\mu,p}^h(\zeta,\eta)} \times \\
&&\hspace{-2cm}
\times \Int_{0}^{1}
dy\frac{y^{n-1/2-ik}}{\sqrt{1-y}} \half\Erfc\left[\sqrt{\frac{k}{2Q^2(K,k)}}\left(\bar\xi_{\mu,p}^{h}(k,\zeta,\eta)-\ln(y)\right)\right]
_{\big|_{\eta=0}}\as k\infty
~,\nonumber
\end{eqnarray}
\eseq
where:
\begin{eqnarray}
\label{B2xi}
\bar{\xi}_{\mu,p}^{h}(k,\zeta,\eta)&\equiv&\bar{L}_{\mu,p}^h(\zeta,\eta)-\frac{k{\cal A}}{2}
~.
\end{eqnarray}
To leading order,  \Eq{Umu2} is given by:
\begin{eqnarray}
\label{U}
U_\mu(k)&\sim&C_\mu e^{-i\pi\smoothN (E)+ikL_\mu}Q_{-\half-ik}[1]\half\Erfc\left[\sqrt{\frac{k}{2Q^2(K,k)}}\xi_\mu(k)\right]~,
\as k\infty~,
\end{eqnarray}
which yields \EQ{Nsmooth_final} when inserted into \EQ{NumSmooth1}. Similarly,  the leading oder of \Eq{Uphmmu2} is given by:
\begin{eqnarray}
\label{Upm}
U_{p,h,m,\mu}(\zeta,k)&\sim&
\sum_{n=0}^{\left[\frac{m}{2}\right]}C_{\mu}^{(p,m)}A_{h,n}^{(p,m)}\left[\frac{1}{\Chp},\partial{\eta},\eta\right]
e^{-i\pi\smoothN (E)+ik\bar{L}_{\mu,p}^h(\zeta,\eta)} \times\nonumber \\
&&
\times B\left[n+\half-ik,\half\right]\half\Erfc\left[\sqrt{\frac{k}{2Q^2(K,k)}}\bar\xi_{\mu,p}^{h}(k,\zeta,\eta)\right]
_{\big|_{\eta=0}}
\Lf&&\hspace{8.5cm}\as k\infty
~.
\end{eqnarray}
In \Eq{Upm} the $\partial_\eta$ differentiations must now be performed. But since the objective here is only to derive a leading order approximation, the derivatives of the complementary error function can be neglected. The reason for this is that the derivatives of  complementary error functions are proportional to Gaussians centered their cut-off. Since their contribution is proportional to their width, The Gaussians' contribution is smaller than the leading term, proportional to the complementary error function. Thus, the leading order approximation of \Eq{Nosc} is \EQ{Nosc_final}:
\begin{eqnarray*}
{\cal N}_{po}&&\big(\zeta,E{(k)}\big)\sim\frac{1}{2\pi} \Im\Bigg\{\sum_{\{p\}}\sum_{h\in \Gamma/\Gamma_{p}}\bar{\Delta}^{(p,h)}(\zeta,k)\Bigg\}
\as k\infty~,
\end{eqnarray*}
where, using \Eq{Ap}:
\beqa\label{delta_ph_def}
\bar{\Delta}^{(p,h)}(\zeta,k)&\equiv&-i\sum_{m=0}^{\infty}\sum_{n=0}^{\left[\frac{m}{2}\right]}\frac{(2n-1)!!}{(2n)!!}B\left[n+\half-ik,\half\right] \times\nonumber\\
&&\times 
a_{m-2n}\left[\frac{1}{\Chp},ik-\half-2n\right] \left(\frac{1}{\Chp}\right)^{\half+2n}\times\nonumber\\
&&\times
\sum_\mu C_{\mu}^{(p,m)}e^{-i\pi\smoothN (E)+ik\bar{L}_{\mu,p}^h(\zeta,0)}\Erfc\left[\sqrt{\frac{k}{2Q^2(K,k)}}\xi_{\mu,p}^{h}(k,\zeta)\right]
~.
\eeqa
In the last formula, the definition:    $\xi_{\mu,p}^{h}(k,\zeta)\equiv\bar\xi_{\mu,p}^{h}(k,\zeta,\eta=0)$ was used.

Using arguments similar to those leading from \EQ{delta_i_oded} to \Eq{delta_i_ophir}, \EQ{delta_ph_def} can be simplified to yield \EQ{delta_ph} (for the details see \App{appn2}):
\beqar
\label{delta_ph_appn}
\bar{\Delta}^{(p,h)}(\zeta,k)&=&\sum_\mu \Delta^{(p,h)}_{\mu}(\zeta,k) e^{-i\pi\smoothN (E)+ikL_{\mu,p}}\Erfc\left[\sqrt{\frac{k}{2Q^2(K,k)}}\xi_{\mu,p}^{h}(k,\zeta)\right]
~,
\eeqar
where:
\beqa\label{Delta_phmu_0}
\Delta^{(p,h)}_{\mu}(\zeta,k)&=&-i\sum_{j=0}^{j_{p}}\bar{C}_\mu^{(p,j)}e^{-ik(j+1)L_p}Q_{-\half-ik}\left[\cosh{d_{hp^{j+1}h^{-1}}(\zeta)}\right]
~.
\eeqa
The $\bar{C}_\mu^{(p,j)}$'s were defined in \Eq{def_Cmupj} and the definitions of \Eq{neg_u_p_maslov} for even boosts were used.

In order to be consistent the leading asymptotic behavior of the functions appearing in \EQ{Delta_phmu_0} must be used. Since (BV G.21):
\beqar
Q_{-\half-ik}\left[\cosh{d}\right] \sim \sqrt{\frac{i\pi}{2k\sinh{d}}}\e^{ikd} \as{k}\infty
~,
\eeqar 
one obtains \EQ{Delta_phmu_approx}.

	\setcounter{equation}{0}\def\theequation{\Alph{section}.\arabic{equation}}
\section{The derivation of Eq.~\protect\ref{delta_ph_appn}}\label{appn2}
In this appendix, \EQ{delta_ph_appn} is derived from \EQ{delta_ph_def}. Since $\bar{\Delta}^{(p,h)}(\zeta)$ involves only absolutely convergent sums, the summation order can be exchanged and the sum over $\mu$ performed last. Using the definition of $\xi_{\mu,p}^{h}(k,\zeta)$ (after \EQ{delta_ph}) and the definition of $A_{h,n}^{(p,m)}$ \Eq{Ap} in \EQ{delta_ph_def}, one obtains:
\beqa\label{delta_ph_1}
\bar{\Delta}^{(p,h)}(\zeta,k)&=&\sum_\mu \Delta^{(p,h)}_{\mu}(\zeta,k) e^{-i\pi\smoothN (E)+ikL_{\mu,p}}\Erfc\left[\sqrt{\frac{k}{2Q^2(K,k)}}\xi_{\mu,p}^{h}(k,\zeta)\right]
~,
\eeqa
where
\beqar
\Delta^{(p,h)}_{\mu}(\zeta,k)&\equiv&-i
\sum_{m=0}^{\infty}C_{\mu}^{(p,m)}\sum_{n=0}^{\left[\frac{m}{2}\right]}\frac{(2n-1)!!}{(2n)!!}B\left[n+\half-ik,\half\right]
\times\nonumber\\&&\times 
a_{m-2n}\left[\frac{1}{\Chp},ik-\half-2n\right] \left(\frac{1}{\Chp}\right)^{\half+2n-ik}
~.
\eeqar
Using the definition of the $C_{\mu}^{(p,m)}$'s (see \EQ{Cmupm}) yields:
\beqa
\label{E2}
\Delta^{(p,h)}_{\mu}(\zeta,k)&=&-iC_{\mu-p}(-1)^{j_{p}}e^{-\half(j_{p}+1)L_{p}}\sum_{j=0}^{j_{p}}(-1)^jd_{j_{p}-j}\left(e^{-L_{p}}\right)\times\Upsilon_j\left[L_p,k\right]
\eeqa
where:
\beqar
\Upsilon_j\left[L_p,k\right]&\equiv&
\sum_{m=0}^{\infty}e^{-m(j+1)L_{p}}\sum_{n=0}^{\left[\frac{m}{2}\right]}\frac{(2n-1)!!}{(2n)!!}B\left[n+\half-ik,\half\right]\times\nonumber\\&&\times 
a_{m-2n}\left[\frac{1}{\Chp},ik-1/2-2n\right] \left(\frac{1}{\Chp}\right)^{\half+2n-ik}
~.
\eeqar
Exchanging the sum over $m$ with the sum over $n$ and shifting summation (opposite to the operation in \EQ{sumExchange}) yields an expression similar to the one that appears in \Eq{legendre-g}, that led us to \Eq{legendre-g2}. Therefore:
\beqar
\Upsilon_j\left[L_p,k\right]=e^{-ik(j+1)L_{p}}e^{\half(j+1)L_{p}}Q_{-\half-ik}\left[\cosh{d_{hp^{j+1}h^-1}(\zeta)}\right]
~.
\eeqar
Insertion of the last expression into \EQ{E2} and \ref{delta_ph_1} with the definitions of \EQ{def_Cmupj} and \ref{neg_u_p_maslov}, leads to \EQ{delta_ph_appn}.

\bibliographystyle{unsrt}

\newpage
\pagestyle{empty}
\vspace*{7cm}
\centering{\Large \bf Figure captions:}
\def\thefigure{\arabic{figure}}
\setcounter{figure}{0}
\begin{figure}[p]
\caption{{\bf Left} - The action of some isometries of the group generated by reflections at the sides of the shaded triangle on it. One can see that the copies tile the pseudo-sphere. {\bf Right} - The billiard chosen for the numerical calculations (a magnification of the shadowed area on the left).}
\label{tile}
\end{figure}
\begin{figure}[p]
\caption{$d^{\perp}_{g}(\protect\zeta)$ is the distance between a point $\protect\zeta$ and the invariant geodesic of $g$ measured along a geodesic that is perpendicular to it. The invariant geodesic of $g$ is aligned along the $y$ axis.}
\label{fig_perp}
\end{figure}
\begin{figure}[p]
\caption{Comparison between the zeros of $\Delta(E)$ approximated by (\protect\ref{Delta_summed}) and the exact eigenenergies (empty circles) found numerically.}
\label{fig_spec}
\end{figure}
\begin{figure}[p]
\caption{The integration contours used for resummation.}
\label{contour}
\end{figure}
\begin{figure}[p]
\caption{Comparison of an exact cut through the ground-state ($E_0\approx118.9$) density (thin line) with the semiclassical result, before including periodic orbits (dashed line) and after including them (thick line). On the right is the exact ground-state density in the whole billiard, with the cut through it denoted by a line.}
\label{wavecut1}
\end{figure}
\begin{figure}[p]
\caption{Same as Fig. \protect\ref{wavecut1} but for the first excited state ($E_1\approx201.1$).}
\label{wavecut2}
\end{figure}
\begin{figure}[p]
\caption{The scar weight on the four shortest primitive periodic orbits ($p=1$ denotes the shortest,  $p=2$ the next shortest, etc.) calculated on the zeros of the approximate spectral determinant. The vertical lines are the locations of the eigenenergies predicted by the Bohr-Sommerfeld quantization rule (\protect\ref{mybohr}).}
\label{figYp1234}
\end{figure}
\begin{figure}[p]
\caption{The four shortest periodic orbits: solid line - the shortest orbit, dashed - next in length, dot-dashed - longer still and the fourth shortest is dotted.}
\label{figShorts}
\end{figure}
\begin{figure}[p]
\caption{(a) - Density of eigenfunction \#$84$ ($k=67.0$) and the second shortest periodic orbit. The fourth shortest periodic orbit on the density of eigenfunction \#$68$ ($k=60.8$) (b), \#$90$ ($k=69.1$) (c) and  \#$91$ ($k=69.5$) (d).}
\label{waves84689091}
\end{figure}
\begin{figure}[p]
\caption{Density of eigenfunction \#$92$ ($k=69.9$) (a), and density of eigenfunction \#$93$ ($k=70.2$) (b). On both figures the shortest periodic orbit is presented, as well as a tube of width $\protect\frac{1}{\protect\sqrt{k}}$ around it.}
\label{waves9293}
\end{figure}
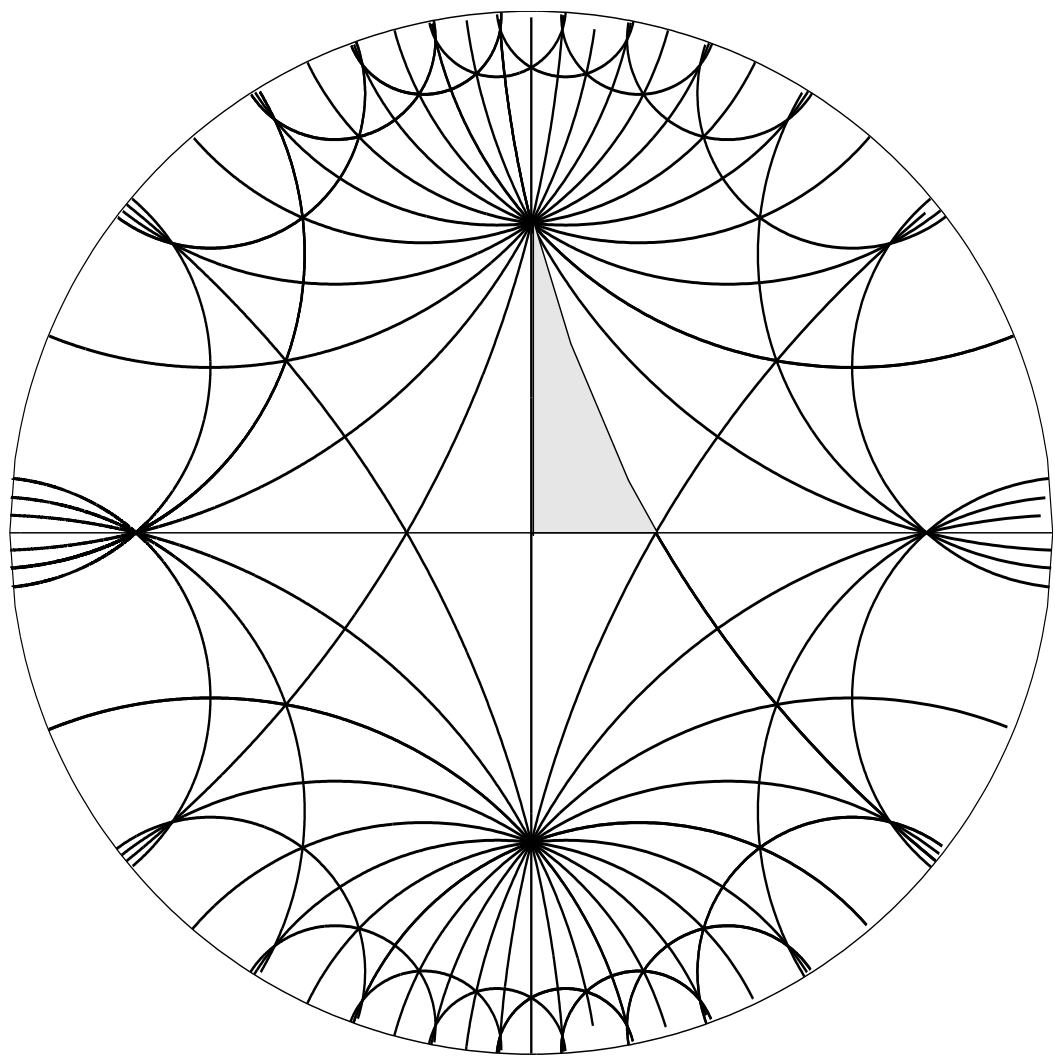
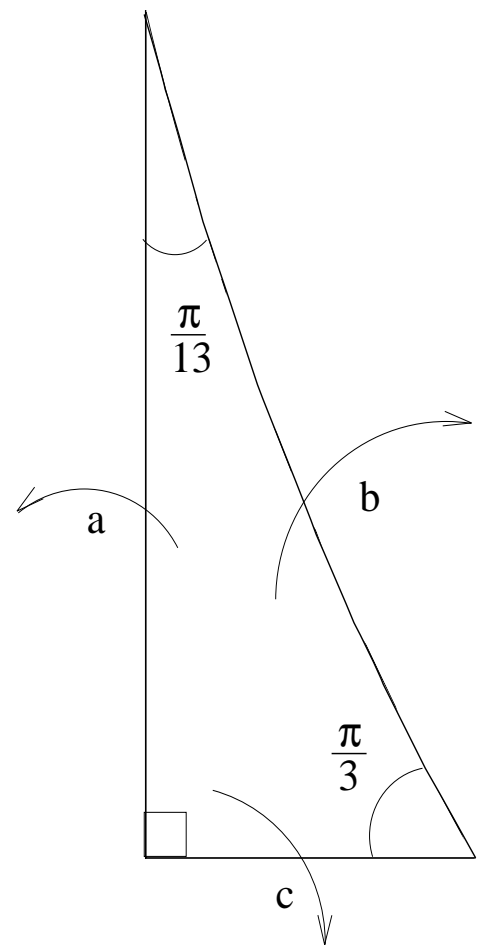
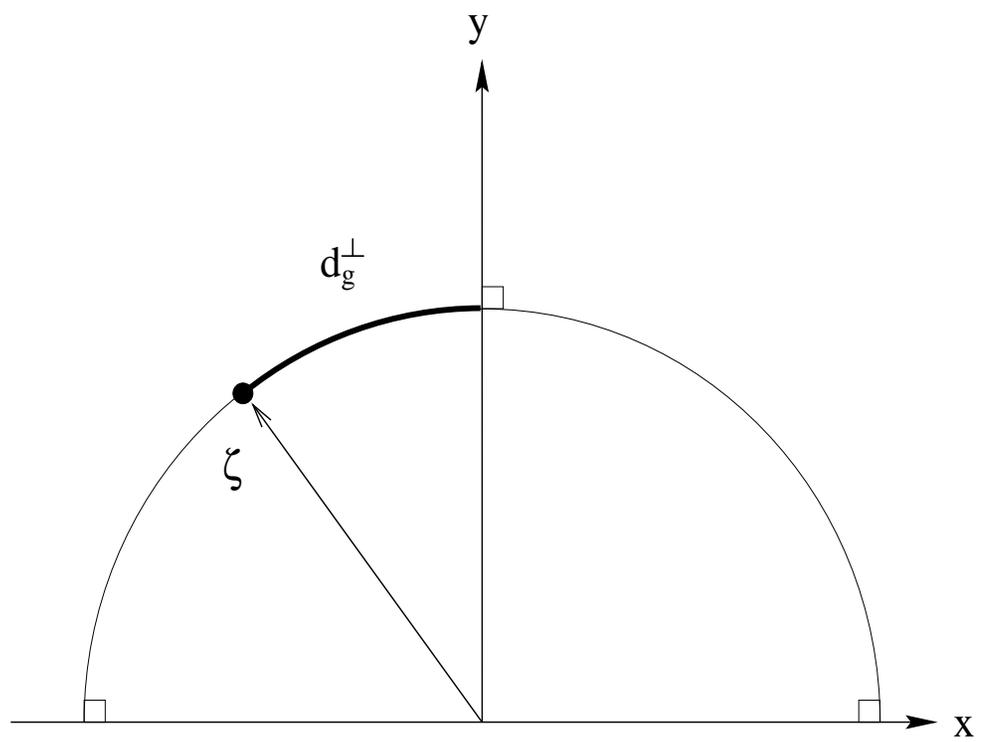
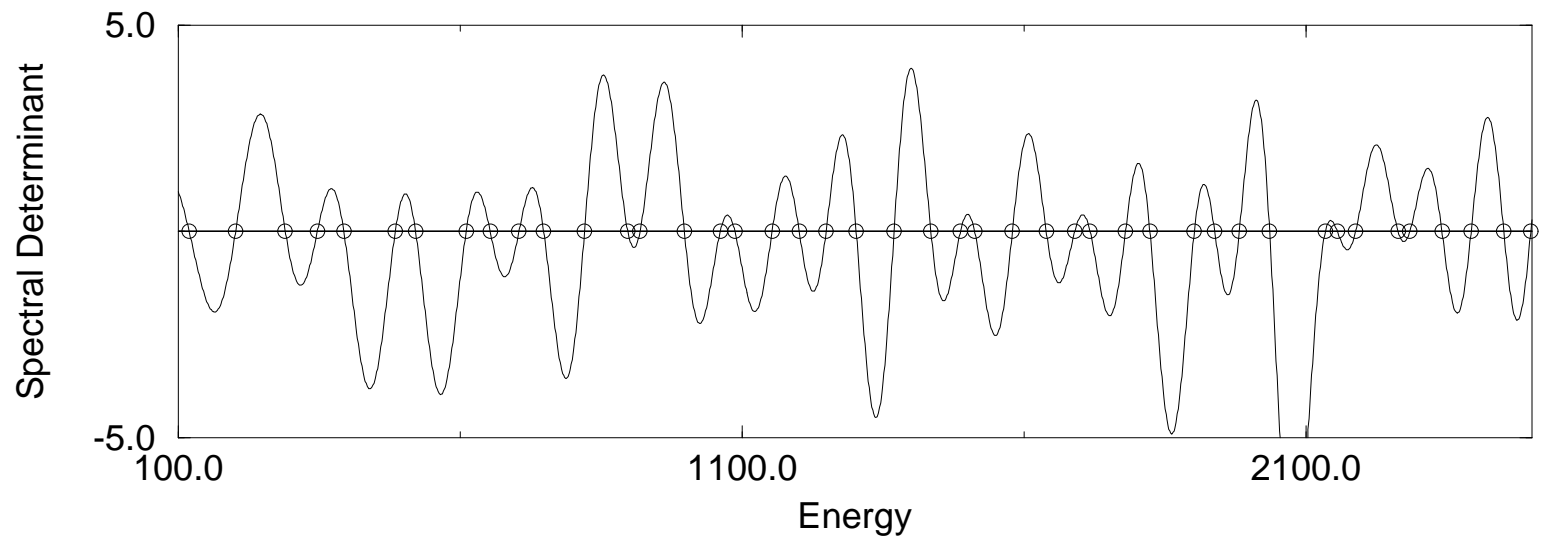
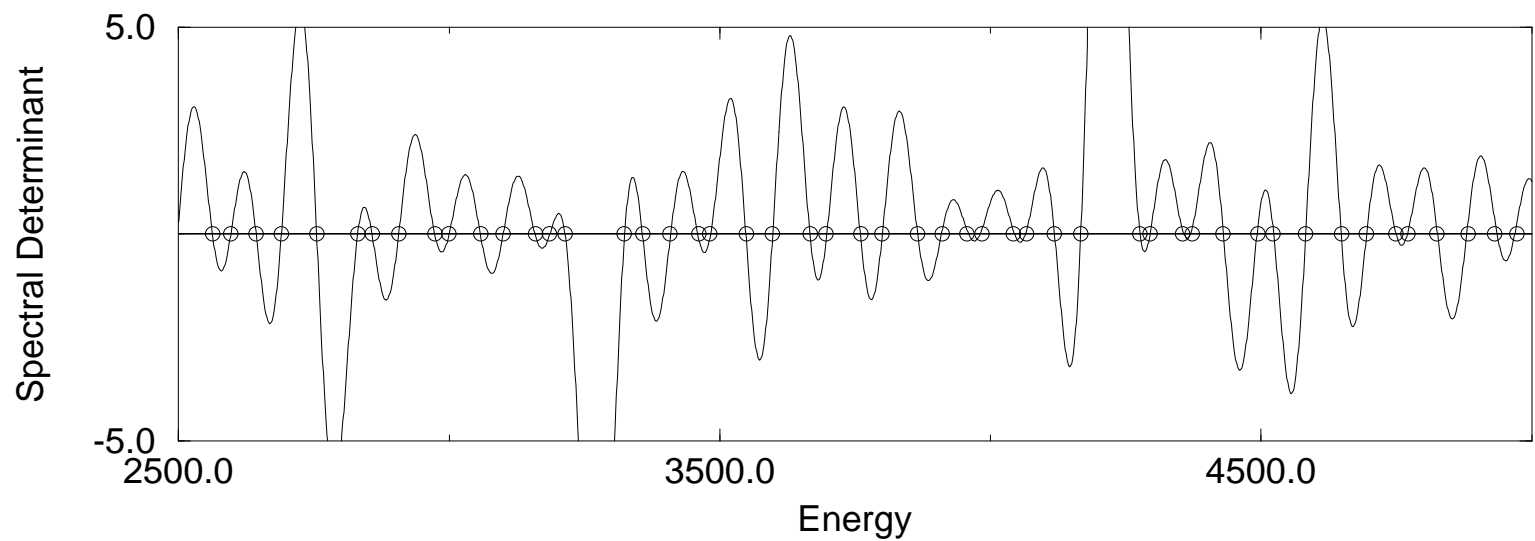
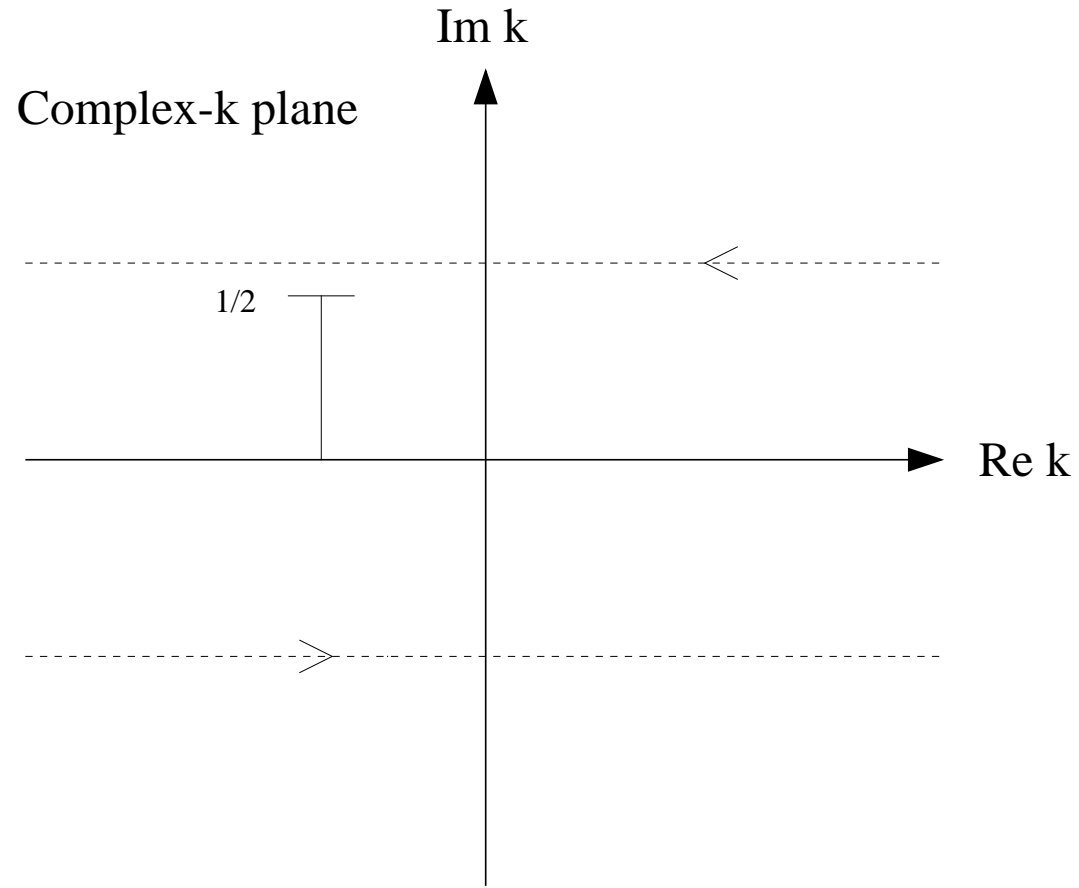
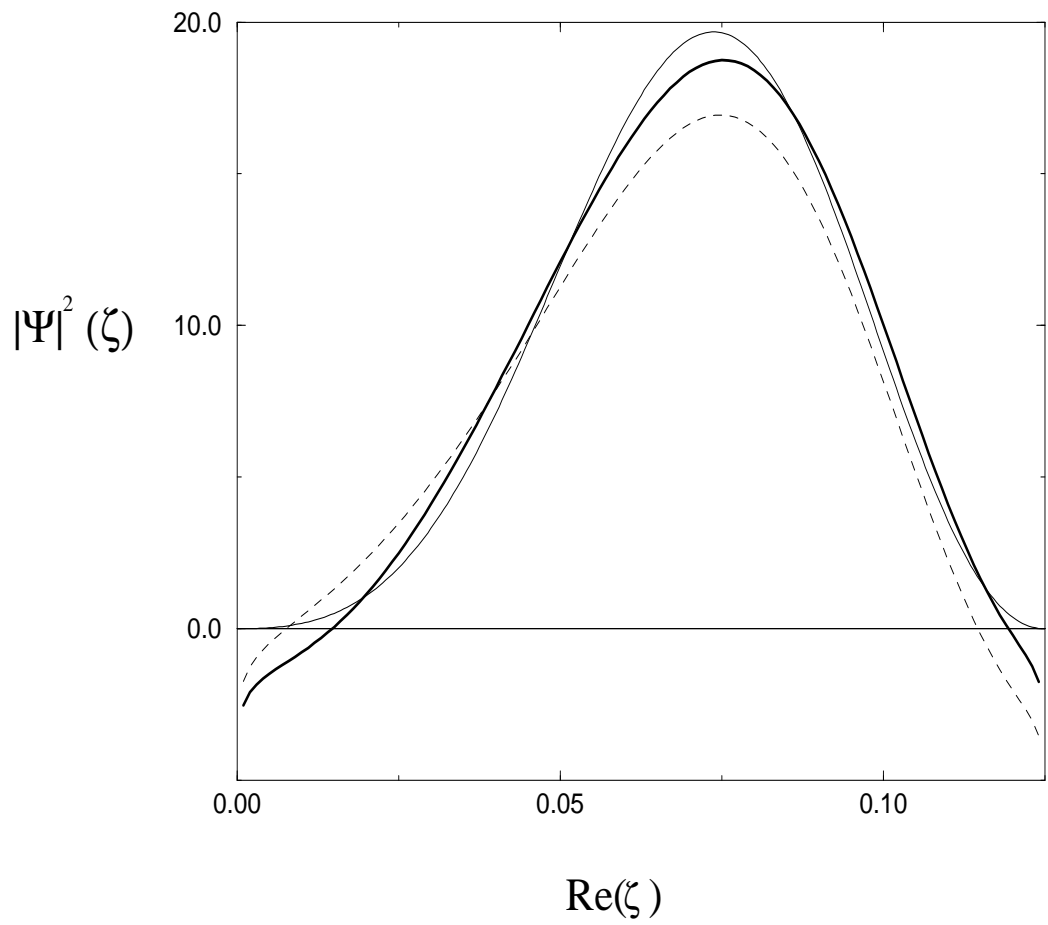
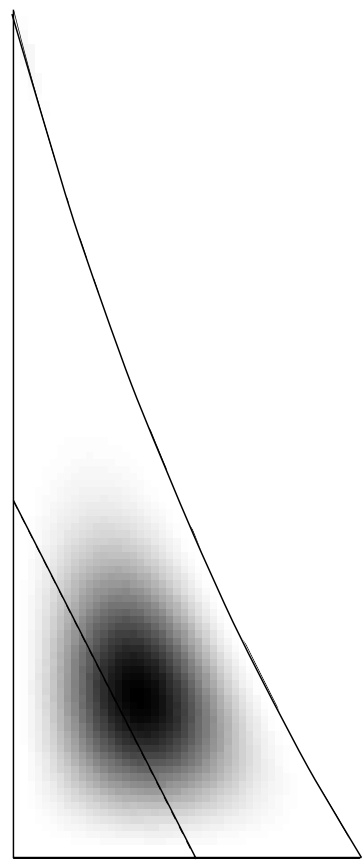
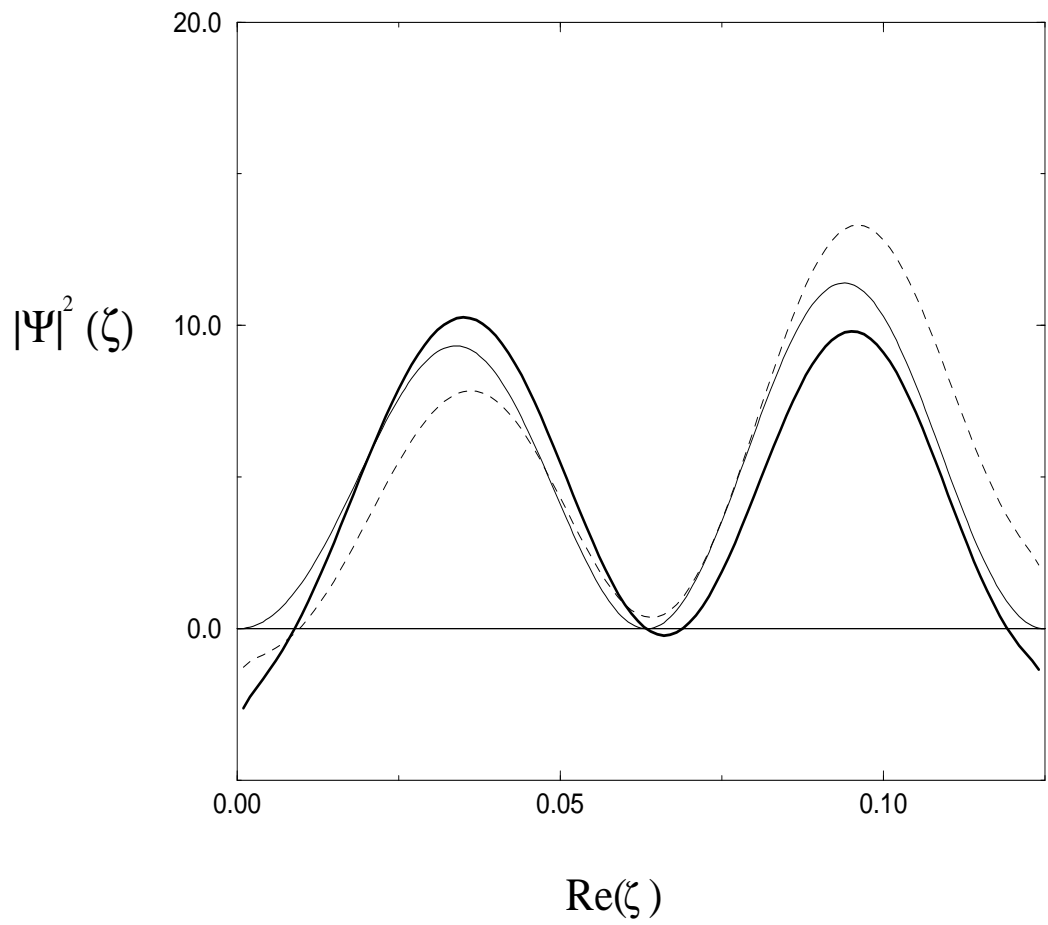
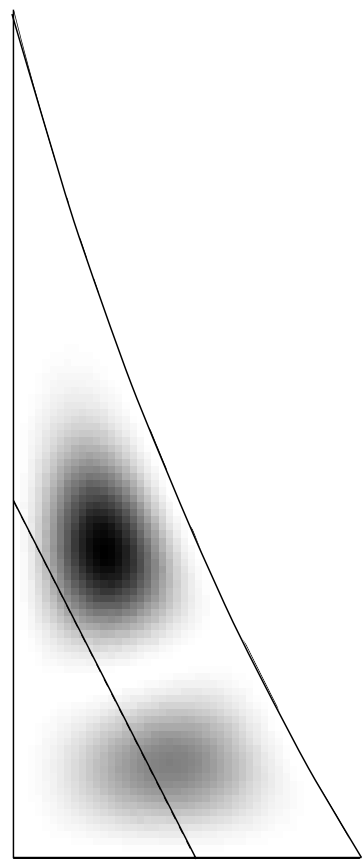
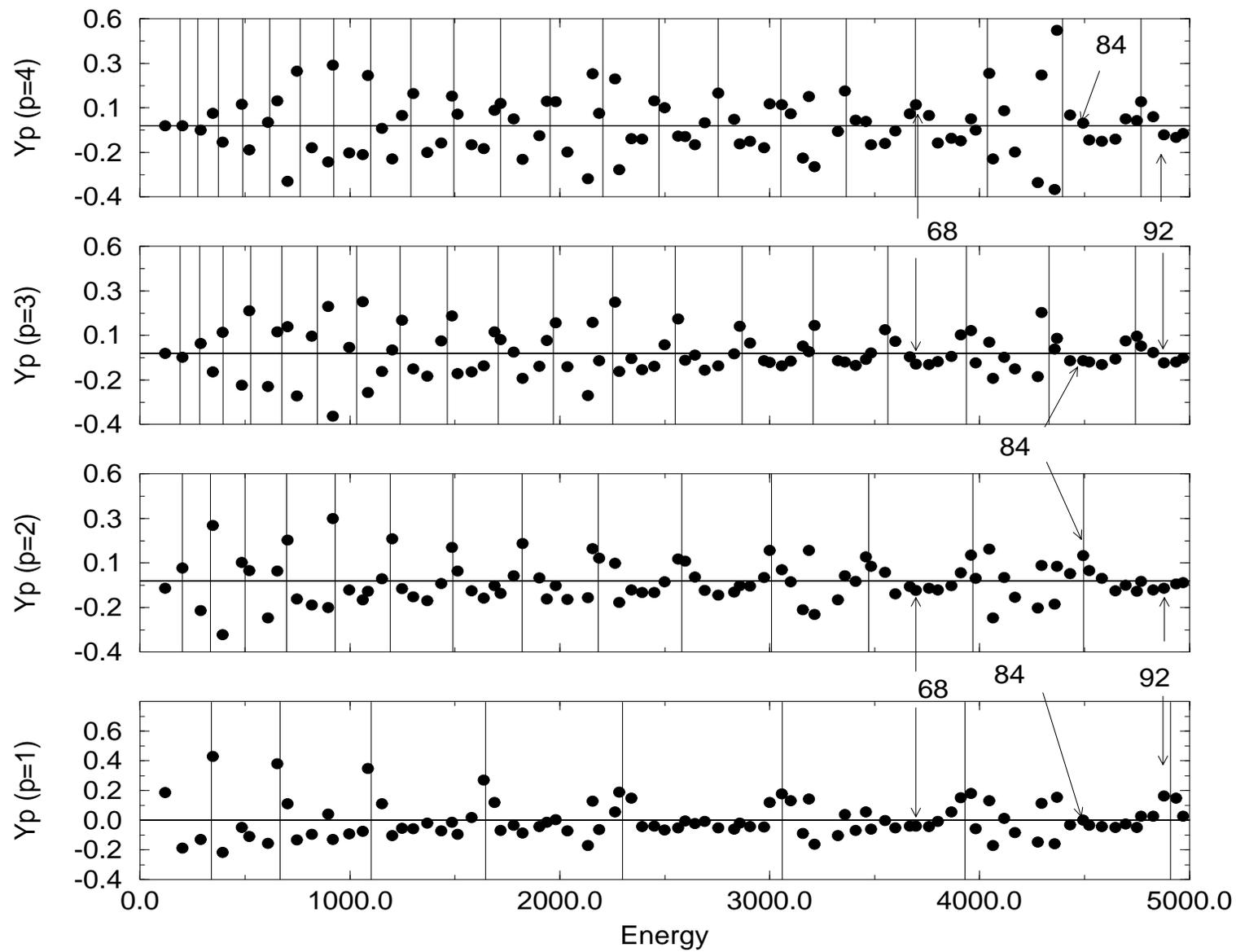
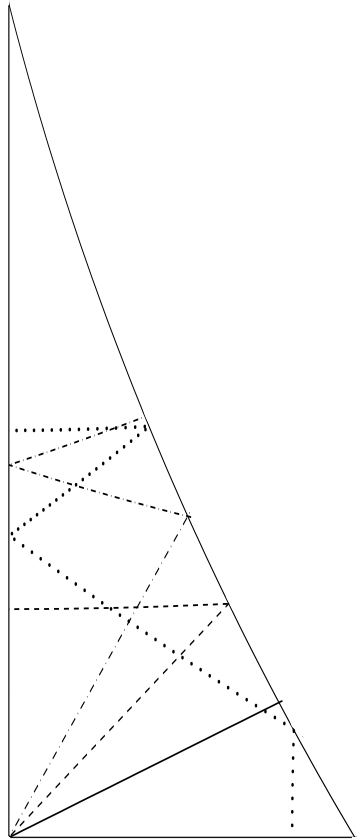
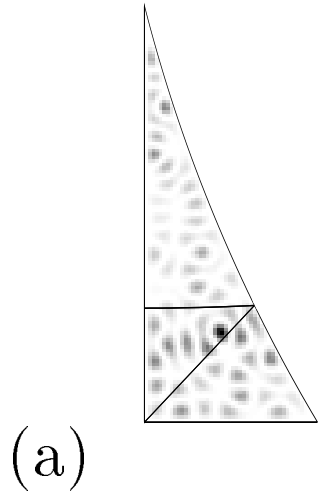
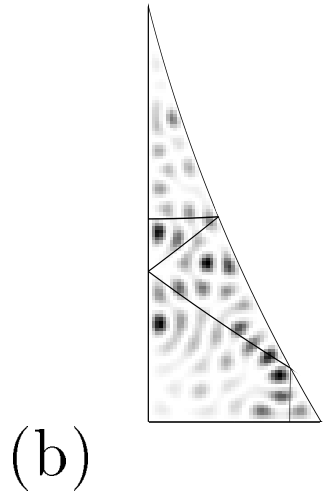
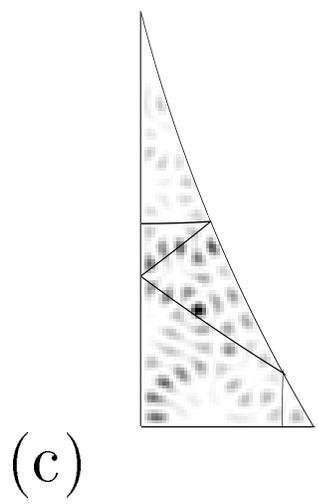
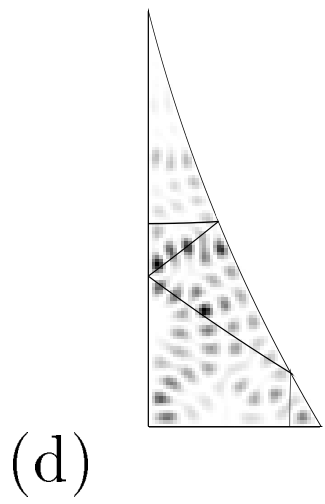
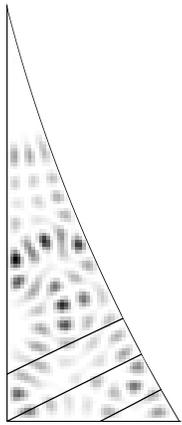
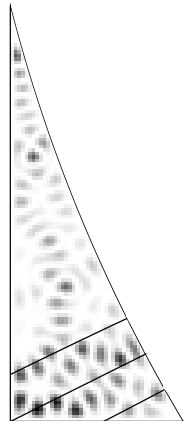
\end{document}